\documentclass[structabstract]{aa}
\usepackage{xspace}
\usepackage{graphicx}
\usepackage{natbib}
\usepackage{amssymb}
\usepackage{longtable}
\usepackage{caption}
\usepackage{subfig}
\usepackage{supertabular}
\usepackage{lscape}
\usepackage{rotating}
\newcommand{\lsim}{\ensuremath{\,\lesssim\,}\xspace}
\newcommand{\K}{\ensuremath{\,{\rm K}}\xspace}

\newcommand{\degree}{\ensuremath{\,^\circ}\xspace}

\newcommand{\hii}{{\rm H\,{\footnotesize II}}\xspace}
\newcommand{\hiismall}{{\rm H\,{\scriptsize II}}\xspace}

\newcommand\micron{\mbox{$\mu$m}}
\newcommand\nodata{ ~$\cdots$~ }

\begin{document}

\title{Distinguishing between H{\Large II} regions and planetary
  nebulae with Hi-GAL, WISE, MIPSGAL, and GLIMPSE\thanks{{\it
      Herschel} is an ESA space observatory with science instruments
    provided by European-led Principal Investigator consortia and with
    important participation from NASA.}}
\titlerunning{Distinguishing \hiismall\ regions from PNe}
\authorrunning{Anderson et al.}

\author{L.~D.~Anderson\inst{1, 5} \and A.~Zavagno\inst{1} \and
  M.~J.~Barlow\inst{2} \and
  P.~Garc\'ia-Lario\inst{3} \and
  A.~Noriega-Crespo\inst{4}}

\institute{Laboratoire d'Astrophysique de Marseille (UMR 6110 CNRS \& Universit\'e de Provence), 38 rue F. Joliot-Curie, 13388 Marseille Cedex 13, France. \and
Department of Physics and Astronomy, University College London, Gower Street, London WC1E 6BT, UK \and
Herschel Science Centre, European Space Astronomy Centre, ESA, Madrid, Spain \and
SPITZER Science Center, California Institute of Technology, Pasadena, CA 91125 \and
Current Address: Physics Department, West Virginia University,
Morgantown, WV 26506, USA.\\Loren.Anderson@mail.wvu.edu}

\date{Received / Accepted}

\abstract{\hiismall\ regions and planetary nebulae (PNe) both emit at
  radio and infrared (IR) wavelengths, and angularly small
  \hiismall\ regions can be mistaken for PNe.  This problem of
  classification is most severe for \hiismall\ regions in an early
  evolutionary stage, those that are extremely distant, or those that
  are both young and distant.  Previous work has shown that
  \hiismall\ regions and PNe can be separated based on their infrared
  colors.}  {Using data from the $Herschel$ Hi-GAL survey, as well as
  WISE and the $Spitzer$ MIPSGAL and GLIMPSE surveys, we wish to
  establish characteristic IR colors that can be used to distinguish
  between \hiismall\ regions and PNe.}  {We perform aperture
  photometry measurements for a sample of 126 \hiismall\ regions and
  43 PNe at wavelengths from 8.0\,\micron\ to 500\,\micron.}  {We find
  that \hiismall\ regions and PNe have distinct IR colors. The most
  robust discriminating color criteria are $[F_{12}/F_{8}] < 0.3$,
  $[F_{160}/F_{12}] > 1.3$, and $[F_{160}/F_{24}] > 0.8$ (or
  alternately $[F_{160}/F_{22}] > 0.8$), where the brackets indicate
  the log of the flux ratio.  All three of these criteria are
  individually satisfied by over 98\% of our sample of
  \hiismall\ regions and by $\sim10\%$ of our sample of PNe.
  Combinations of these colors are more robust in separating the two
  populations; for example all \hii\ regions and no PNe satisfy
  $[F_{12}/F_{8}] < 0.4$ and $[F_{160}/F_{22}] > 0.8$.  When
    applied to objects of unknown classification, these criteria prove
    useful in separating the two populations.  The dispersion in
  color is relatively small for \hiismall\ regions; this suggests that
  any evolution in these colors with time for \hiismall\ regions must
  be relatively modest.  The spectral energy distributions (SEDs) of
  \hiismall\ regions can be separated into ``warm'' and ``cold''
  components.  The ``cold'' component is well-fit by a grey-body of
  temperature 25\,K.  The SEDs of nearly two-thirds of our sample of
  \hiismall\ regions peak at 160\,\micron\ and one third peak at
  70\,\micron.  For PNe, 67\% of the SEDs peak at 70\,\micron, 23\%
  peak at either 22\,\micron\ or 24\,\micron, and 9\% (two sources)
  peak at 160\,\micron.}{}

\maketitle

\section{Introduction}
Despite their different origins, \hii\ regions and
planetary nebulae (PNe) have similar characteristics at infrared (IR)
and radio wavelengths\footnote{Technically PNe and \hiismall\ regions
  are both ``\hiismall\ regions'' in that they contain ionized plasma.
  Throughout we exclusively refer to ``\hiismall\ regions'' as the
  ionized zones surrounding massive stars, and their associated
  dust and gas.}.  The dust associated with \hii\ regions and PNe is
responsible for their emission at mid-IR (MIR) to far-IR (FIR)
wavelengths.  For \hii\ regions this emission is from dust within the
\hii\ region itself and from the photodissociation region (PDR), which
is the interface between the ionized gas and the interstellar medium
\citep[see recent results from, e.g.][]{deharveng10, anderson10a, rodon10, anderson11b}.
The IR emission from the PDRs of \hii\ regions comes from neutral
material possibly collected during the expansion of the \hii\ region.
For PNe, the IR emission is from the dusty layers of material shed during
their creation.

Recently, there has been extensive work on the IR photometric
properties of large samples of PNe.  PNe have been studied using the
IRAC instrument \citep{fazio04} on {\it Spitzer} by \citet{hora04,
  cohen07b, kwok08, phillips08b, zhang09, cohen11} and also with the
MIPS instrument \citep{rieke04} on {\it Spitzer} by \citet{chu09,
  zhang09, phillips11}.  Most pertinent for the present study are the
results of \citet{cohen07b, cohen11}, who showed that PNe should be
well-separated from both ultra-compact (UC) \hii\ regions and also
more evolved \hii\ regions in IRAC MIR color-color diagrams.  These
results, however, were based on simulated \hii\ region colors and
it remains to be seen if they can be replicated in real measurements.



\hii\ regions and PNe also emit at radio wavelengths.  In both cases,
the central star(s) produce ultra-violet radiation that ionizes the
surrounding medium.  This ionized gas emits radio continuum radiation
via thermal Bremsstrahlung (free-free).  The radio continuum emission
from \hii\ regions is extremely bright; it has been known for some
time that most bright continuum sources in the Galactic plane are
\hii\ regions.  For example, 75\% of the bright radio continuum
sources in \citet{westerhout58} are \hii\ regions and none are PNe.
While not as bright, the radio continuum emission from PNe has also
been detected for numerous objects \citep{milne75, zijlstra89,
  aaquist90, condon98b}.  The stars powering \hii\ regions emit
considerably more ionizing flux than those powering PNe, and therefore
the radio emission from \hii\ regions can be detected at larger
distances.


Numerous authors have used data from the IRAS point source catalog to
establish criteria for identifying \hii\ regions and PNe.
\citet{wc89b} found that UC \hii\ regions, which are generally young
and small in angular size, have characteristic colors of
$[F_{25}/F_{12}] \ge 0.57$ and $[F_{60}/F_{12}] \ge 1.30$, where
$F_\lambda$ is the IRAS flux at wavelength $\lambda$ and the brackets
indicate a base ten logarithm.  \citet{hughes89} found that
\hii\ regions have characteristic colors $[F_{25}/F_{12}] \ge 0.4$ and
$[F_{60}/F_{12}] \ge 0.25$ while PNe have $[F_{25}/F_{12}] \ge 0.4$
and $[F_{60}/F_{12}] \le 0.25$.  Using different IRAS flux bands,
\citet{pottasch88} found PNe generally have colors $[F_{25}/F_{12}]
\ge 0.46$ and $[F_{60}/F_{25}] \le 0.52$.  These studies demonstrate
the power of FIR criteria for locating and discriminating between
\hii\ regions and PNe.


Candidate \hii\ regions and PNe can in many cases be distinguished
based on visual inspection of their IR emission
\citep[e.g.][]{cohen07b, cohen11}.  \hii\ regions tend to have more
extended emission that PNe, and they show less symmetry.  The central
stars of PNe are frequently detected at near-IR wavelengths, and this
has been used in the ``Red MSX Survey'' (RMS) survey
\citep{urquhart08} to distinguish between the two types of object.
While a visual inspection of sources is sufficient in many cases,
there are two cases in which visual inspection fails: young
\hii\ regions that are not yet extended at IR wavelengths, and evolved
\hii\ regions at extreme distances from the Sun.  In both cases, the
resolution of current MIR surveys does not allow one to distinguish
between \hii\ regions and PNe.  The problem is the most severe in the
case of young, distant \hii\ regions.  Visual inspection of course
also fails for studies of extragalactic \hii\ regions and PNe where the
resolution is insufficient.

The recently completed Green Bank Telescope \hii\ Region Discovery
Survey \citep[GBT~HRDS][]{bania10, anderson11a} illustrates the
problem of separating \hii\ regions from PNe.  The HRDS sources were
selected on the basis of spatially coincident radio and MIR emission.
While this criterion easily distinguishes thermally emitting sources
(e.g. \hii\ regions and PNe) from non-thermally emitting sources
(e.g. supernova remnants and active galactic nuclei), it is not
sufficient to distinguish between \hii\ regions and PNe.  The number
of PNe detected in the HRDS is likely low \citep{bania10}, although
\citet{anderson11a} mention a number of cases where the source
classification is unclear.  The HRDS sources are the most distant
\hii\ regions yet detected (L.~D.~Anderson, 2011, in prep.).  Their
small angular size may prevent the detection of more diffuse emission,
making visual classification unreliable.

The goal of the present work is to establish IR criteria that can be
used to distinguish between \hii\ regions and PNe.  These criteria can
then be applied to samples of objects of unknown classification, which
could greatly increase the number of known \hii\ regions and PNe.  The
{\it Herschel} Hi-GAL survey \citep{molinari10}, together with WISE
\citep{wright10}, {\it Spitzer} MIPSGAL \citep{carey09}, and {\it
  Spitzer} GLIMPSE \citep{benjamin03, churchwell09} allow us to revisit IR criteria
for separating the two classes of objects, at much higher resolution
and sensitivity compared to IRAS.

\section{Data}
\subsection{{\it Herschel} Hi-GAL}
The on-going Hi-GAL survey uses the PACS \citep{poglitsch10} and SPIRE
\citep{griffin10} instruments of the {\it Herschel Space Observatory}
to map the Galactic plane over the zone $+70\degr \ge l \ge -70\degr$,
$|b| \le 1\degr$, and was recently awarded time to map
$\sim240$\,square degrees in the outer Galaxy as well.  Hi-GAL
contains photometry bands centered at 70\,\micron\ and
160\,\micron\ with PACS, and 250\,\micron, 350\,\micron, and
500\,\micron\ with SPIRE.  The spatial resolutions of these bands are
$6.7\arcsec$, $11\arcsec$, $18\arcsec$, $25\arcsec$, and $37\arcsec$,
respectively.  The point source sensitivities of Hi-GAL measured in a
complex field are 0.5, 4.1, 4.1, 3.2, and 2.5 Jy/beam for the
70\,\micron, 160\,\micron, 250\,\micron, 350\,\micron, and
500\,\micron\ bands while those in a less complex field are 0.06, 0.9,
0.7, 0.7, and 0.8 Jy/beam \citep{molinari10}. The FIR coverage of Hi-GAL
traces the emission of dust; the 70\,\micron\ Hi-GAL band is sensitive
to warm ($\sim100$ \,K) dust while the longer wavelength bands are
more sensitive to colder $(\sim10-30)$\,K dust.

\subsection{WISE}
The Wide-field Infrared Survey Explorer ($WISE$) mapped the entire sky
in four IR bands: 3.4\,\micron, 4.6\,\micron, 12\,\micron, and
22\,\micron.  The data used here are from the preliminary data release
of 14 April 2011, which cover $\sim57\%$ of the sky.  Along the
Galactic plane, this coverage extends approximately from $60\degr
\ge\ \ell \ge -70\degree$ and $240\degr \ge \ell \ge 120\degree$.  The
spatial resolutions in the four bands are $6\farcs1$, $6\farcs4$,
$6\farcs5$, and $12\arcsec$ and the sensitivities are 0.08\,mJy,
0.11\,mJy, 1\,mJy, and 6\,mJy, respectively.

Here we use only the WISE 12\,\micron\ and 22\,\micron\ bands.  The
WISE 12\,\micron\ band should trace similar dust emission components
as that of the IRAC~8.0\,\micron\ band.  The 12\,\micron\ band,
however, is significantly broader than that of the IRAC~8.0\,\micron\
band; within this broad band are PAH features at 11.2\,\micron,
12.7\,\micron, and 16.4\,\micron\ \citep[see][]{tielens08}. The PAH
features at 7.7\,\micron\ and 8.6\,\micron\ also fall within the
bandpass although at diminished sensitivity.  There are also nebular
emission lines that may be strong in evolved PNe.  The
22\,\micron\ band should trace the same dust emission components as
the 24\,\micron\ MIPSGAL band (see below).

The WISE 22\,\micron\ band saturates for point sources at 12.4\,Jy
(see WISE explanatory
supplement\footnote{http://wise2.ipac.caltech.edu/docs/release/prelim/expsup/\\wise\_prelrel\_toc.html})
which is six times higher than that for MIPSGAL, 2\,Jy
\citep{carey08}.  We thus expect the WISE 22\,\micron\ band to be less
affected by problems caused by saturation compared to the MIPSGAL
24\,\micron\ band.  The WISE 12\,\micron\ band saturates for point
sources at 0.9\,Jy.

The WISE image data have units of DN and we use the DN-to-Jy
conversion factors of $2.9045\times10^{-6}$ and $5.2269\times10^{-5}$
for the 12\,\micron\ and 22\,\micron\ bands, respectively (see WISE
explanatory supplement$^2$).

\subsection{{\it Spitzer} MIPSGAL}
The {\it Spitzer} MIPSGAL survey mapped the Galactic plane over
$+60\degr \ge l \ge -60\degr$, $|b| \le 1\degr$ using the MIPS
instrument.  MIPSGAL used two of the three MIPS photometric bands
centered at 24\,\micron\ and 70\,\micron.  Here we use only data from
the 24\,\micron\ band, which has a spatial resolution of $6\arcsec$.
MIPSGAL saturates at $\sim1700$\,MJy\,sr$^{-1}$ \citet{carey08}, and
strong saturation is present for many star-forming regions and for
some PNe (see \S\ref{sec:results}).  Unlike the IRAC bands, the MIPS
24\,\micron\ data have minimal correction for extended source fluxes
\citep{cohen09} and we apply no such correction when computing fluxes.

For \hii\ regions, the 24\,\micron\ emission is detected from two
distinct zones: along the PDR and also spatially coincident with the
ionized gas.  These components contribute roughly equally to the
24\,\micron\ flux of \hii\ regions \citep{deharveng10, anderson11b}.
The emission coincident with the ionized gas likely comes from very
small grains (VSGs) out of thermal equilibrium.  The emission from the
PDR is probably a combination of emission from VSGs and from a larger
grain population whose temperature is closer to 25\,\K.  For PNe, the
emission at 24\,\micron\ also likely contains contributions from dust
at a range of temperatures, but there may also be nebular emission
lines that contribute to the emission in this band.

\subsection{{\it Spitzer} GLIMPSE}
The {\it Spitzer} Galactic Legacy Infrared Mid-Plane Survey
Extraordinaire (GLIMPSE) mapped the Galactic plane over the same
Galactic zone as MIPSGAL using the IRAC instrument.  GLIMPSE contains
four MIR bands at 3.6\,\micron, 4.5\,\micron, 5.8\,\micron, and
8.0\,\micron\ at resolutions of $\sim2\arcsec$.  Here we use only the
8.0\,\micron\ band because, relative to the other GLIMPSE bands, 
this band contains fewer stars and more intense diffuse
emission.  This makes source fluxes easier to measure and decreases
errors in the aperture photometry.  In addition to being sensitive to
warm dust, the 8.0\,\micron\ filter contains polycyclic aromatic
hydrocarbon (PAH) emission from PAH features at 7.7\,\micron\ and
8.6\,\micron\ \citep[see][]{tielens08}.  PAH molecules fluoresce when
excited by far-UV photons and thus emit very strongly in \hii\ regions
\citep[see {\it ISO} spectra in][]{peeters02b, peeters02a}.  PAH
emission is moderately strong for C-rich PNe and is weak or absent for
O-rich PNe \citep{volk03, bernard-salas06}.

Scattering within the focal plane causes an increase in the measured
flux of extended sources with the IRAC instrument, an effect that is
wavelength dependent.  To correct for this effect, we follow the {\it
  Spitzer} recommendations\footnote{See
  http://web.ipac.caltech.edu/staff/jarrett/irac/\\calibration/index.html}
and apply an aperture correction to the 8.0\,\micron\ fluxes, based on
the size of the aperture.  This correction lowers the measured flux
values by a maximum of 35\% for apertures $\gtrsim 50\arcsec$.

\subsection{IRAS}
The {\it IRAS} satellite mapped 98\% of the sky at 12, 25, 60, and
100\,\micron.  In \S\,\ref{sec:compare} we use reprocessed IRAS data
known as IRIS \citep{miville05} to compare against
WISE~12\,\micron\ fluxes.  IRIS improves on the original IRAS data by
correcting for calibration issues, the zero level, and striping
problems.  The angular resolution of the IRIS data is
$3\farcm8\pm0\farcm2$, $3\farcm8\pm0\farcm2$, $4\farcm0\pm0\farcm2$,
and $4\farcm3\pm0\farcm2$ for the 12, 25, 60, and 100\,\micron\ bands,
respectively.

\begin{figure*}
\includegraphics{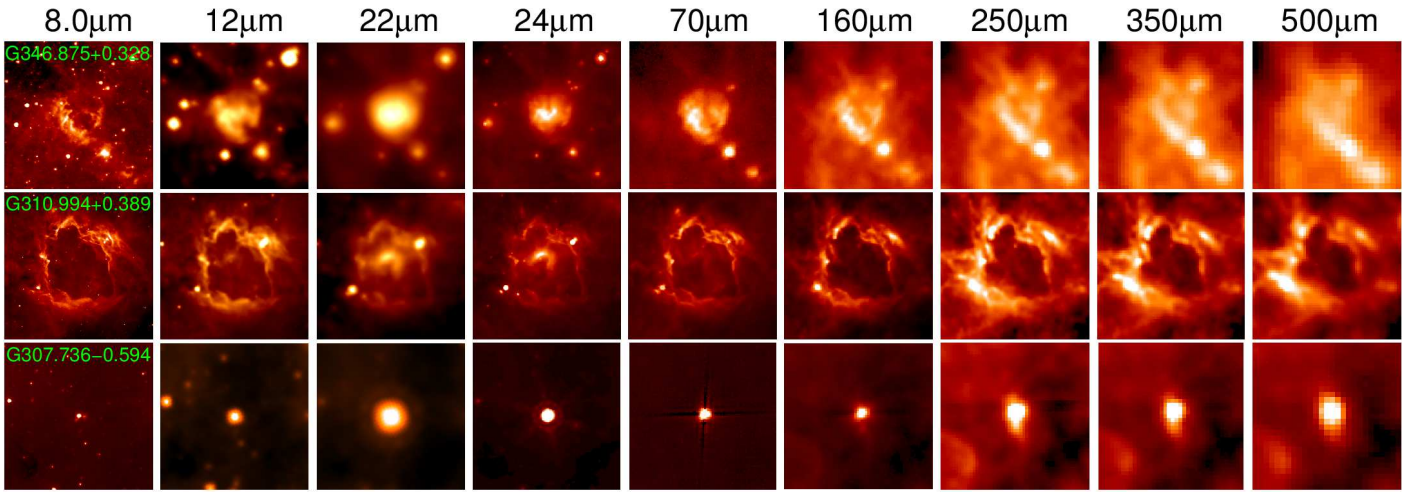}

\caption{Infrared images of \hiismall\ regions from 8.0\,\micron\ to
  500\,\micron\ for three representative sources.  Shown are
  G346.875+0.328, G310.994+0.389, and G307.736$-$0.594 from top to
  bottom.  The images for G346.875+0.328 and G307.736$-$0.594 are
  $5\arcmin$ on a side while those of G310.994+0.389 are $12\arcmin$
  on a side.}
\label{fig:hii}
\end{figure*}

\section{Source Samples}
We require that all \hii\ regions and PNe fall within the boundaries
of all the IR surveys used here.  At the time of this writing, the
Hi-GAL coverage is the most restrictive.  It includes longitude
coverage from $346\fdg5 > \ell > 290\degree$, and $3\fdg5 > \ell >
-4\fdg5$, in addition to a few other more isolated fields.  We
restrict our samples to the above stated ranges in the southern
Galactic plane and about the Galactic center.

\subsection{H{\tiny II} Regions}
Our goal in creating the \hii\ region sample is to have \hii\ regions
spanning a range of evolutionary stages.  We make no attempt to create
an \hii\ region sample with angular sizes matching that of the PN
sample.  While many \hii\ regions could not be confused with PNe due
to their large angular sizes, if they were located 20\,kpc from the
Sun the situation would be less clear.  By including \hii\ regions
spanning a range of physical sizes, we can account for the two
situations where an \hii\ region may be confused with a PN: young
compact \hii\ regions in an early evolutionary stage and more evolved
\hii\ regions at extreme distances from the Sun.

We compile our sample of \hii\ regions from (a) the HRDS
\citep{anderson11a}, (b) the catalog of \hii\ regions detected in
recombination line emission prior to the HRDS described in
\citet{anderson11a} (henceforth identified as the ``known'' sample),
and (c) sources classified as \hii\ regions in the RMS survey.  The
angular sizes of the HRDS targets are typically $\lsim 2\arcmin$, but
many are extremely distant (L.~D.~Anderson, 2011, in prep.).  The
regions known prior to the HRDS consist mainly of those sources in the
recombination line surveys of \citet{caswell87} and \citet{lockman89}
for the Galactic zone here studied, although there are also
contributions from \citet{lockman96}, and \citet{sewilo04}.  The RMS
survey contains objects detected as point sources in the MSX survey
\citep{egan03}, excluding the zone within $10\degree$ of the Galactic
center.  Since they are unresolved by MSX, which has an angular
resolution of 18\farcs3 \citep{price01}, the MSX targets are small in
angular size.  The MSX sources used here that were classified as
\hii\ regions by the RMS survey all have detected CO emission
\citep{urquhart07b} and radio continuum emission \citep{urquhart07a},
which together solidify their classification as \hii\ regions.  As
with the PNe, we only include \hii\ regions in relatively
uncomplicated fields.

Our \hii\ region sample contains 126 sources: 21 from the HRDS, 49
from the known sample, and 56 from the RMS survey.  This sample of
\hii\ regions contains older, evolved regions in the ``known'' sample,
younger \hii\ regions from the RMS survey (many of which are in the
ultra-compact phase, as indicated by their compact radio continuum
emission measured by \citet{urquhart07a}), and the HRDS sources, which
likely span a range of evolutionary stages (L.D. Anderson, 2011, in
prep.).

Example \hii\ region images are shown in Figure~\ref{fig:hii}, which shows
data from 8.0\,\micron\ to 500\,\micron.  At 8.0\,\micron\ and 12\,\micron, if the
resolution is sufficient, the \hii\ region PDR appears bright
in part because of emission of PAHs.
At 24\,\micron\ and 70\,\micron, if the resolution is sufficient, the
\hii\ region PDRs are detected with a similar morphology as that seen
at 8.0\,\micron.  There is also, however, IR emission from the region
spatially coincident with the ionized gas \citep[see][]{watson08,
  watson09, deharveng10, anderson11a}.  This emission likely comes
from a different dust grain population than the emission in the PDR.
It is detected at wavelengths from 12\,\micron\ to 70\,\micron\ (see
G310.994+0.389 in Figure~\ref{fig:hii}).  At wavelengths longer than
70\,\micron, the dust emission associated with the \hii\ regions here
studied is almost entirely from the colder dust in the PDR.  This
emission again has a similar morphology as the emission at
8.0\,\micron, although nearby local filaments associated with the
\hii\ regions are also sometimes detected.  For angularly small
sources, the emission from the \hii\ region is frequently found along
a filament whose emission at {\it Herschel} SPIRE wavelengths suggests
a cold temperature (for example, the emission from a filament with a
dust temperature of 15\,K peaks near 200\,\micron\ and this filament
would emit strongly in the 250\,\micron\ SPIRE band).  For these
sources, their small angular size and location along a cold filament
together hint at an early evolutionary stage.

In Figure~\ref{fig:hii}, G346.875+0.328 and G310.994+0.389 would not
be confused for PNe due to their extended emission.  For
G307.736$-$0.594, however, the situation is less clear from a visual
inspection alone.

\begin{figure*}
\includegraphics{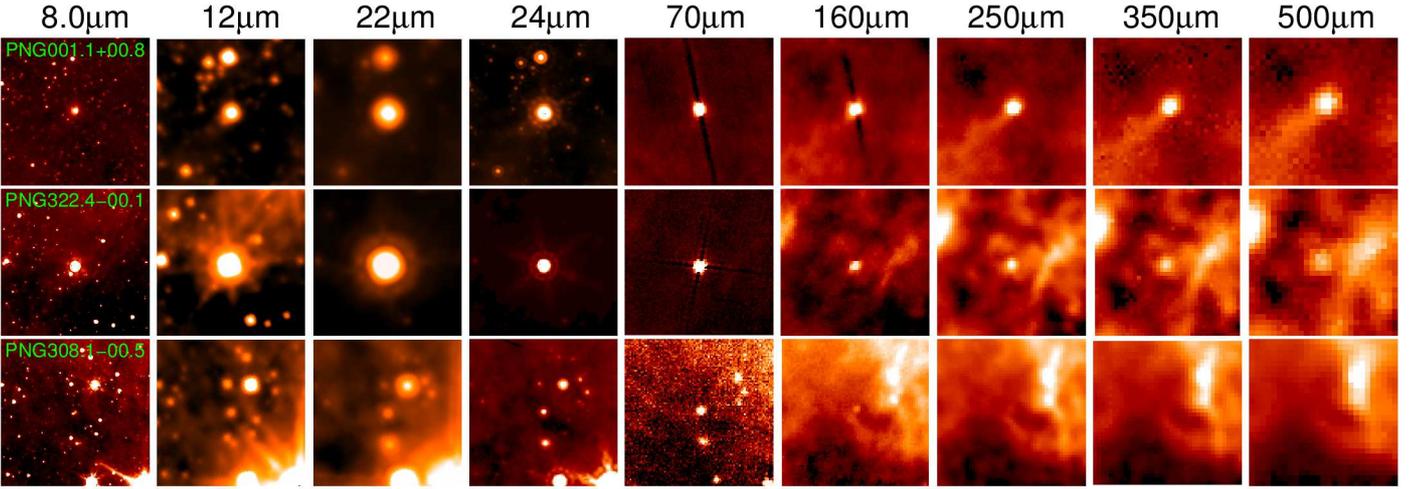}

\caption{Infrared images of PNe from 8.0\,\micron\ to
  500\,\micron\ for three representative sources.  Shown are
  PNG001.1+00.8, PNG322.4$-$00.1, and PNG308.1$-$00.5 from top to
  bottom.  All images are $5\arcmin$ on a side.  PNG001.1+00.8 and
  PNG322.4$-$00.1 are rather bright sources and are detected at all
  wavelengths, while PNG308.1$-$00.5 is rather faint and is not
  detected at wavelengths longer than 160\,\micron.  The increasing
  confusion with wavelength is evident.}

\label{fig:pne}
\end{figure*}

\subsection{Planetary Nebulae}
Our goal in creating the PNe sample is to have as many sources as
possible, while excluding sources misclassified as PNe.  Because we
are interested in determining representative FIR colors, we require
that all PNe are detected in at least two of the IR bands considered
here.

We compile our sample of PNe from \citet{kohoutek01}, and the
Macquarie-AAO-Strasbourg H$\alpha$ PN Project (MASH) catalogs of more
recently discovered PNe \citep{parker06, miszalski08}.  We also
include PNG313.3+00.3 from \citet{cohen05}.  Based on their
photometric and spectroscopic observations, the MASH catalogs have a
qualitative estimate of whether each source is a ``true'' PN, or
whether it is merely ``likely'' or ``possible''.  We include only
``true'' PNe in our sample and exclude MASH sources classified as
``likely'' or ``possible'' PNe.  We also exclude PNe in complicated
zones of the Galaxy where accurate photometry would be difficult.
Following \citet{cohen11}, we remove objects in \citet{kohoutek01}
whose classification as PNe is questionable.  Our sample includes 43
PNe, 25 from \citet{kohoutek01}, nine from \citet{parker06}, eight
from \citet{miszalski08}, and one source from \citet{cohen05}.

Example PNe images are shown in Figure~\ref{fig:pne}, which shows data from
8.0\,\micron\ to 500\,\micron.  While many PNe are resolved at
8.0\,\micron, few show extended emission at longer wavelengths.  At
the longest {\it Herschel} wavelengths, the emission from PNe is
frequently confused with brighter diffuse features.

 Our sample of PNe may not be representative of the entire
  population of PNe in the Galaxy.  Our sample is, however,
  representative of IR-bright PNe that may be confused with
  \hii\ regions.  Many PNe are too faint in the IR to be detected with
  the suvey data used here.  For example, \citet{hora08} found that of
  233 optically identified PNe in the Large Magellanic Cloud, 161 were
  detected at 24\,\micron\ with the MIPS instrument.  All sources in
  our sample are found along the Galactic plane, which may introduce a
  bias as it will exclude some of the nearest brightest sources that
  are off the plane.

\section{Aperture Photometry\label{sec:apphot}}
We perform aperture photometry for all PNe and \hii\ regions in our
sample using the Kang software\footnote{http://www.bu.edu/iar/kang/},
version 1.3.  This software treats fractional pixels such that if an
aperture encloses 30\% of the area of a pixel, 30\% of the pixel value
is included in the aperture flux.  Additionally, one can use apertures
of arbitrary size and shape and define regions to exclude from the
aperture photometry.  This flexibility is important for the present
analysis in which many regions of interest are in complicated fields,
have irregular morphologies, or have nearby contaminating sources of
emission.

For the \hii\ regions in our sample, we generally use the same source
and background apertures at all wavelengths.  When defining apertures
we attempt to include all the associated emission at all wavelengths.
Most of the \hii\ regions in our sample are bright at IR wavelengths,
are relatively unconfused with nearby sources of emission, and have a
similar morphology and angular extent at all wavelengths.  For nearly
all \hii\ regions, a single source aperture for all wavelengths is
sufficient to enclose the associated IR flux from the \hii\ region.
We make an exception for some angularly small \hii\ regions and for
regions in areas whose background emission varies strongly with
wavelength.  Very small sources sometimes appear within a small
cluster of similar sources at 8.0\,\micron, although at longer
wavelengths a single source dominates the emission.  An example of
this phenomenon is shown in the third row of Figure~\ref{fig:hii}.
For such sources we locate the 8.0\,\micron\ source that is spatially
coincident with the bright source detected at longer wavelengths.  We
define a small aperture at 8.0\,\micron\ to isolate the emission from
the source responsible for the longer wavelength emission.

We define for each PN separate source and background apertures at each
wavelength.  Due to their small angular sizes and relatively low IR
fluxes, the aperture photometry for the PNe is considerably more
sensitive to the aperture size and placement compared to that for
\hii\ regions.  The choice to define individual apertures at each
wavelength was necessitated in part by the background, which varies strongly
with wavelength.  An appropriate background at 24\,\micron\ for example may
contain large bright diffuse structures at longer wavelengths.

Bright point sources can affect the aperture photometry measurements.
For the wavelengths considered here, point source contamination is
especially strong from 8.0\,\micron\ to 24\,\micron, where the
contaminants are mostly stellar.  We exclude the pixel
locations of especially bright point sources by hand from the aperture
photometry measurements by defining apertures that avoid such features,
but we do not remove point sources from the images.
The \hii\ regions in our sample are in general much brighter
than the combined flux of any point sources within the aperture at the
wavelengths considered here; the removal of point sources would have minimal
impact on the derived fluxes.  Regions of bright extended emission
have a dearth of detected point sources because of the increased
Poisson ($\sqrt{N}$) noise \citep[see][their Figure~2]{robitaille08}.
Removing the flux from all detected point sources would therefore
assign more flux per unit area to the background region compared to
the source aperture, leading to an underestimate of the \hii\ region
flux.  Due to their small angular size, point source contamination is also
not generally a problem for PNe since very few point sources are
spatially coincident with the PNe themselves.  We make no attempt to
exclude all point sources from the background apertures, although
we do define apertures to exclude bright point sources.  Any
flux from low-intensity point sources distributed over the background aperture
is likely present in the source aperture as well.

We perform aperture photometry using the relation:
\begin{equation}
S_\nu = S_{\nu, 0} - \frac{B_\nu}{N_B} \times N_S\,,
\label{eq:flux}
\end{equation}
where $S_\nu$ is the source flux after the background correction,
$S_{\nu, 0}$ is the flux within the source aperture (before background
correction), $B_\nu$ is the flux within the background aperture, $N_B$
is the number of pixels in the background aperture, and $N_S$ is the
number of pixels in the source aperture.  This treatment subtracts the
mean flux within the background aperture from each pixel in the source
aperture.  The use of the mean flux in the background aperture rather
than the median flux should better estimate the contribution from
low-intensity point sources.

For all \hii\ regions and PNe we define four background apertures and
a single source aperture.  We use for the measured flux at a given
wavelength the average of the four background-subtracted fluxes found
after applying Equation~\ref{eq:flux} individually to the four
source-background pairs.  For the uncertainty in flux we use the
standard deviation of the four background-subtracted flux
measurements.

This method for calculating the flux and uncertainties estimates the
true uncertainties associated with our aperture photometry method,
which in most cases are dominated by the choice of the background
region.  In particular, our method accurately estimates errors in
fields with strongly varying background emission.  In such fields the
standard deviation of the pixel values in any given background
aperture may be low, but the differences in computed flux between
different background apertures may be high.

%
%

\section{Results\label{sec:results}}
The results of the aperture photometry for the \hii\ regions and PNe
are given in Tables~\ref{tab:hii} and \ref{tab:pn}, respectively.
Both tables contain the source name followed by the source Galactic
longitude and latitude and the angular size (radius) of the aperture used at
24\,\micron.  For the PNe in Table~\ref{tab:pn}, we list both
the Galactic name and the commonly used source identifier.  The size
column in Table~\ref{tab:hii} is given in arcmin. while that in
Table~\ref{tab:pn} is given in arcsec.  The sizes listed are the
``mean radius,'' calculated from the area of a circular aperture
encompassing the same area as the source aperture.  Both tables then
list in successive columns the flux in Jy and the uncertainties at
8.0\,\micron, 12\,\micron, 22\,\micron, 24\,\micron, 70\,\micron,
160\,\micron, 250\,\micron, 350\,\micron, and 500\,\micron.  If more
than 0.1\% of the source pixels are saturated at 12\,\micron, 22\,\micron, or
24\,\micron\ (see below) these fluxes are marked with
the symbol: $\dagger$.  The final column in both tables is the
reference.

\begin{figure}
\resizebox{\hsize}{!}{\includegraphics{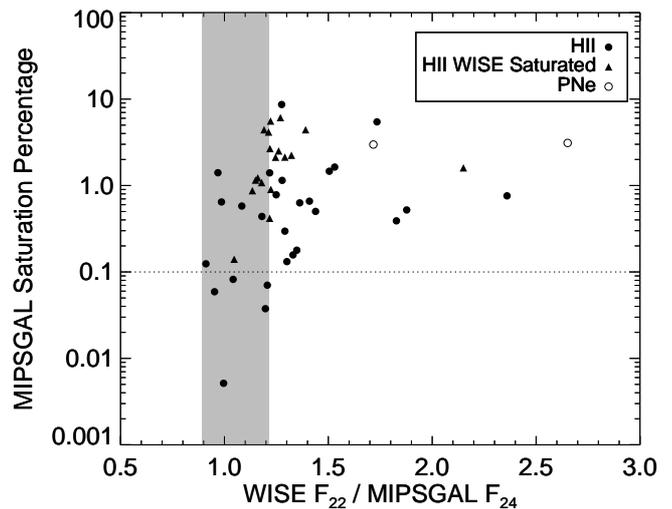}}

\caption{The ratio of the WISE 22\,\micron\ flux to the MIPSGAL
  24\,\micron\ flux versus the percentage of MIPSGAL pixels that are
  saturated.  \hiismall\ regions that have some level of saturation in the
  WISE 22\,\micron\ band are shown with open circles and those that do
  not are shown with filled circles.  The two saturated PNe are shown
  with triangles.  The dotted horizonal line shows our cutoff at 0.1\%
  saturated pixels and the grey area shows the standard
  deviation of the average of the ratio for all non-saturated
  \hiismall\ regions.  The cutoff at 0.1\% separates sources more than one
  standard deviation from the mean.}

\label{fig:mips_saturation}
\end{figure}

If 0.1\% or more the pixels within a source aperture at a given
wavelength are saturated such that they have a value of ``NAN,'' we
exclude this source from all subsequent analyses involving that
wavelength.  When there is such severe saturation the aperture
photometry fluxes are less reliable.  Saturation in the MIPSGAL
24\,\micron\ data is present for many \hii\ regions and for two PNe in
our sample.  The WISE data at 12\,\micron\ and 22\,\micron\ are also
saturated for many sources.  In Figure~\ref{fig:mips_saturation} we
show the ratio of the WISE 22\,\micron\ flux to the MIPSGAL
24\,\micron\ flux versus the percentage of pixels within the MIPSGAL
aperture that are saturated.  The average $F_{22}$ to $F_{24}$ ratio
for the entire sample of \hii\ regions for which no saturation is
detected is $1.05\pm0.16$ (see below); this is shown shaded in
Figure~\ref{fig:mips_saturation}.  The horizontal line at 0.1\%
separates sources whose 22\,\micron\ and 24\,\micron\ fluxes are more
than 25\% different from each other.

Saturation is a problem for many \hii\ regions in the data used here.
Forty \hii\ regions have more than 0.1\% of their pixels at
24\,\micron\ saturated with a value of ``NAN'': 26 from the Known
sample, 12 from the RMS, and two from the HRDS.  Additionally, there
are two PNe that satisfy this criterion: PNG358.9$-$00.7 and
PNG359.3$-$00.9.  There is also strong saturation in the WISE data,
especially at 12\,\micron.  Thirty \hii\ regions in our sample have
more than 0.1\% of their pixels saturated at 12\,\micron\ (25 from the
known sample, four from the RMS and one from the HRDS) and 16 satisfy
this criterion at 22\,\micron\ (15 from the Known sample and one from
the RMS).  In the WISE bands no PNe in our sample are saturated
according to our saturation criterion.  PNG359.3-00.9, a PN that shows
severe saturation at 24\,\micron, however, has an anomalously low flux
at 12\,\micron\ compared to its 8.0\,\micron\ flux.  This indicates
that it too may be saturated at 12\,\micron; we exclude it from
further analysis involving the 12\,\micron\ flux.  Some \hii\ regions
show mild saturation at 250\,\micron\ and 350\,\micron, but none meet
the 0.1\% cutoff.


\addtocounter{table}{1}

\addtocounter{table}{2}

\subsection{Detection Rates\label{sec:detection}}
At each wavelength we visually assess whether each source was
detected.  Some sources that we count as non-detections may in fact be
marginally detected, but they do not allow the high photometric accuracy
required for the present analyses.  The detection of a source at a
particular wavelength is a function of the intrinsic source
brightness, the detector and telescope sensitivity, the manner in
which the data were taken, and the local background.  Whether a source
was detected or not at a particular wavelength therefore does not
necessarily reflect the intrinsic properties of a source.  On average,
however, sources detected at a given wavelength are brighter than
those that are not detected at that wavelength.

All \hii\ regions in our sample are detected at all wavelengths.  All
PNe in our sample are detected at 24\,\micron\ and 70\,\micron.
When creating our PNe sample, however, there were four PNe only
detected in MIPSGAL that otherwise met our criteria; because we
require detection in two wavelength bands, we excluded these sources
from our PNe sample.  This suggests that the 24\,\micron\ MIPSGAL band
is the most sensitive of the bands studied here for locating PNe,
followed by the Hi-GAL 70\,\micron\ band.   \citet{hora08} also
  found that the 24\,\micron\ band was very sensitive to PNe.

All but one PN is detected in the WISE 22\,\micron\ data for a
98\% detection rate while three PNe are not detected in the WISE
12\,\micron\ data for a 93\% detection rate.  At 8.0\,\micron, 91\% of
our sample PNe are detected.  At wavelengths higher than 70\,\micron,
the detection percentage steadily decreases: 72\% of the PNe are
detected at 160\,\micron, 49\% at 250\,\micron, 37\% at 350\,\micron,
and 28\% at 500\,\micron.  In part because of the decreasing spatial
resolution, as the wavelength increases there is considerable
confusion with other sources of emission.  This contributes to the
decreasing detection rates with increasing wavelength, although the
intrinsic faintness of PNe at FIR wavelengths is likely a larger
factor.

\subsection{Fluxes\label{sec:fluxes}}
The \hii\ regions in our sample have significantly higher fluxes
compared to the PNe, at all wavelengths.  Both samples of objects are
composed of relatively bright examples of their classes.  The maximum
PN flux can be used to set an upper limit that potentially would allow
one to separate the two classes.

\begin{figure}
\resizebox{\hsize}{!}{\includegraphics{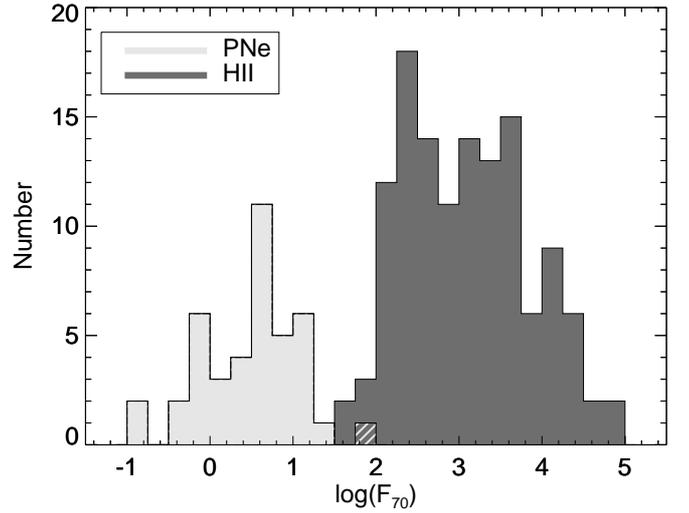}}

\caption{The distributions of 70\,\micron\ fluxes.  The fluxes of
  \hiismall\ regions are shown in dark grey while those of PNe are
  shown in light grey.  The hatched box has contributions from both
  \hiismall\ regions and PNe.  Our sample of \hiismall\ regions is
  significantly brighter than our sample of PNe.}
\label{fig:flux_70}
\end{figure}

In Figure~\ref{fig:flux_70}, we show the two flux distributions for
\hii\ regions and PNe in the 70\,\micron\ Hi-GAL data; the
distributions for other photometry bands are similar.  For our sample
of PNe, the maximum fluxes at 8.0\,\micron, 12\,\micron, 22\,\micron,
24\,\micron, 70\,\micron, 160\,\micron, 250\,\micron, 350\,\micron,
and 500\,\micron\ are 5.3\,Jy, 6.2\,Jy, 83.2\,Jy, 31.4\,Jy, 101.6\,Jy,
46.0\,Jy, 22.3\,Jy, 7.7\,Jy, and 3.5\,Jy, respectively.  The maximum
24\,\micron\ value is saturated and the true value should be
closer to the 22\,\micron\ value.  Objects with fluxes higher
than these values are unlikely to be PNe, although in extreme cases a
bright PNe may have IR fluxes considerably higher than these limits
(e.g. IC\,418 or NGC\,6302).

\subsubsection{\label{sec:compare}A Comparison Between WISE, MIPS, and IRAS}
\begin{figure}
\resizebox{\hsize}{!}{\includegraphics{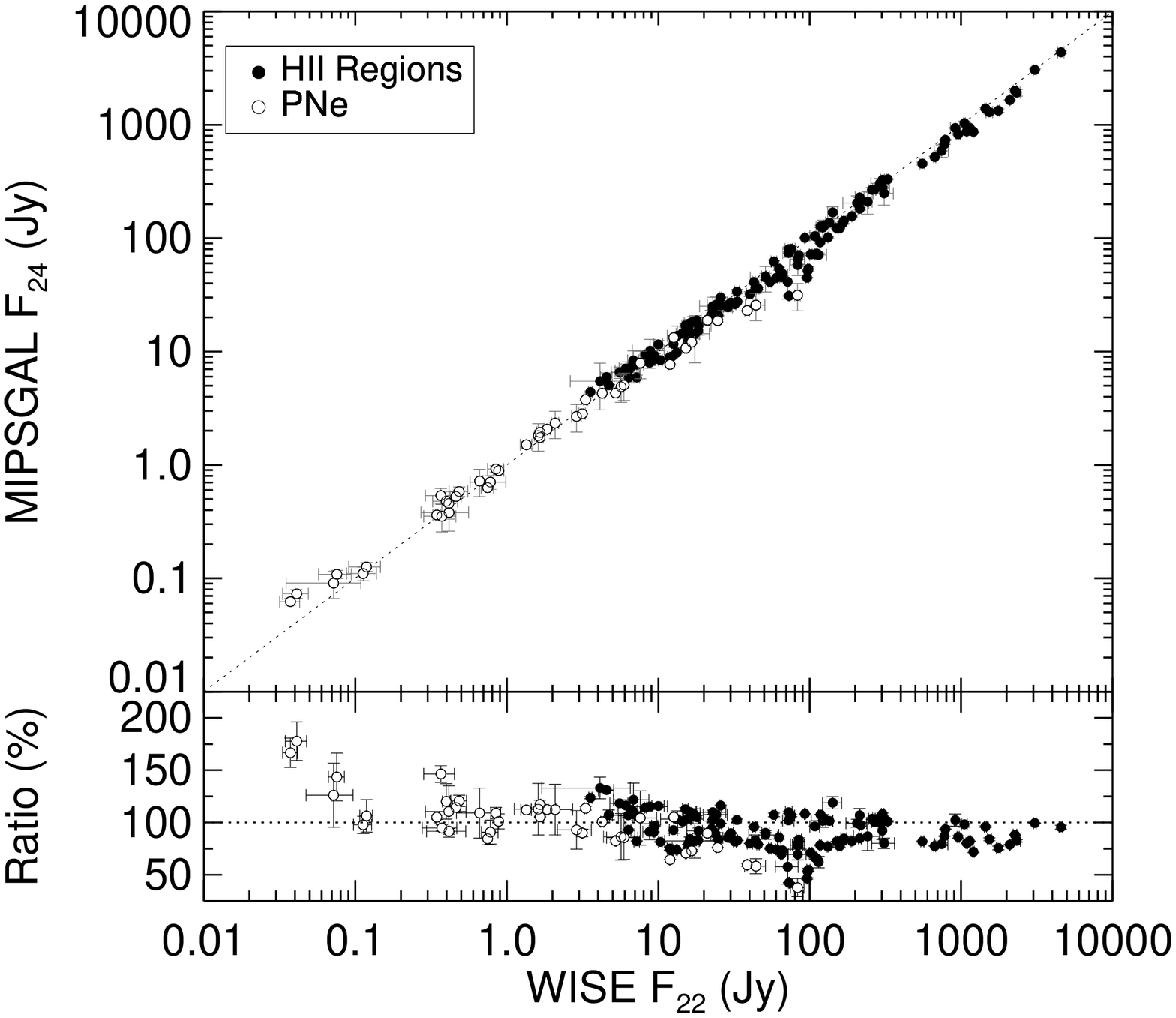}}

\caption{ Correlation between WISE and MIPSGAL.  In the top panel
    we plot the WISE 22\,\micron\ flux versus the MIPSGAL
    24\,\micron\ flux; in the bottom panel we show the ratio of the
    MIPSGAL 24\,\micron\ flux to the WISE 22\,\micron\ flux, as a
    percentage.  The dotted line in both panels shows a one to one
    relationship.  Errors are the $1\sigma$ photometric uncertainties.
    There is a strong correlation between the two fluxes over five
    orders of magnitude.  The sources below the trend line, especially
    for high values of the flux, indicate MIPSGAL saturation.}
\label{fig:mipsgal_wise}
\end{figure}

The 22\,\micron\ WISE fluxes are highly correlated with the
24\micron\ MIPSGAL fluxes, as shown in Figure~\ref{fig:mipsgal_wise}.
This is expected due to their similar wavelengths and bandpass shapes.
That the WISE and MIPSGAL fluxes are so highly correlated over five
orders of magnitude shows that the fidelity of the WISE preliminary
release data is adequate for our analyses here.  We find that the WISE
22\,\micron\ band fluxes are on average $1.05\pm0.16$ times the MIPS
24\,\micron\ fluxes for \hii\ regions that are below our saturation
threshold.  For PNe, the WISE 22\,\micron\ band fluxes are on average
$1.01\pm0.23$ times the MIPS 24\,\micron\ fluxes.  Many of the sources
below the line in Figure~\ref{fig:mipsgal_wise} are saturated in the
24\,\micron\ band -- we have not excluded saturated sources from the
figure.

\begin{figure}
\resizebox{\hsize}{!}{\includegraphics{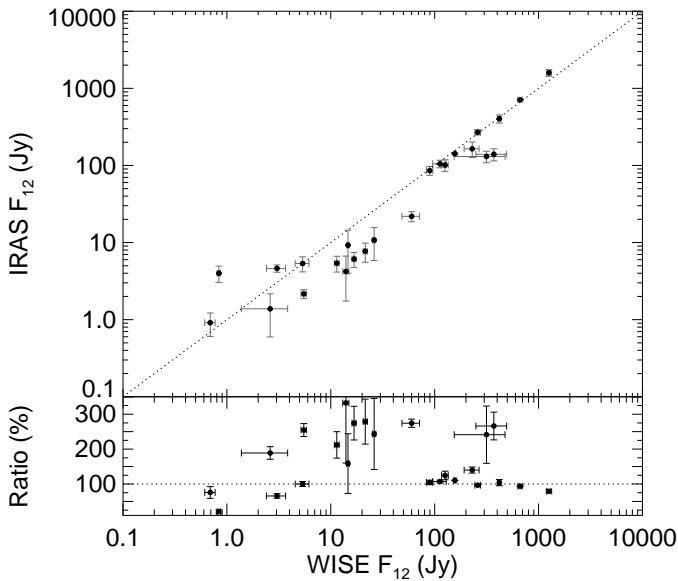}}

\caption{ Correlation between WISE and IRAS.  In the top panel we
    plot the WISE 12\,\micron\ flux versus the IRAS 12\,\micron\ flux;
    in the bottom panel we show the ratio of the IRAS
    12\,\micron\ flux to the WISE 12\,\micron\ flux, as a percentage.
    The dotted line in both panels shows a one to one relationship.
  Errors are the $1\sigma$ photometric uncertainties.  There is a
  correlation between the two fluxes over four orders of magnitude,
  although there is significant scatter in the relationship.  This
  scatter is likely due in large part to the difficulty of IRAS
  photometric measurements.  We have included only \hiismall\ regions
  that are not saturated in the WISE~12\,\micron\ band.}
\label{fig:iras_wise}
\end{figure}

The 12\,\micron\ WISE fluxes are less strongly correlated with IRAS
12\,\micron\ fluxes, as shown in Figure~\ref{fig:iras_wise}.  To
create Figure~\ref{fig:iras_wise} we measured the
IRAS~12\,\micron\ fluxes for well-isolated \hii\ regions.  We used the
same method as before, but we defined new apertures to take into
account the significantly poorer IRAS resolution.  Contrary to
Figure~\ref{fig:mipsgal_wise}, Figure~\ref{fig:iras_wise} shows only
sources that are not saturated in the WISE~12\,\micron\ band according to
our saturation criterion.  Especially at higher fluxes, we find that
the correlation is quite good.  Given the larger photometric errors,
Figure~\ref{fig:iras_wise} is less conclusive than
Figure~\ref{fig:mipsgal_wise}, although it does also suggest that the
photometric quality WISE preliminary release data is high.

\subsection{Spectral Energy Distributions\label{sec:sed}}
Based on the results from IRAS mentioned in the introduction, which
showed that \hii\ regions and PNe possess on average different FIR
colors, we expect that the shapes of the spectral energy distributions
(SEDs) of \hii\ regions and PNe are also different.  Spectral
observations with the {\it Infrared Space Observatory} ({\it ISO\/})
have also shown that PNe and \hii\ regions have different mean SEDs
\citep[see][]{cohen99, peeters02a};  the SEDs of PNe are generally
broader than those of \hii\ regions.

\begin{figure}
\resizebox{\hsize}{!}{\includegraphics{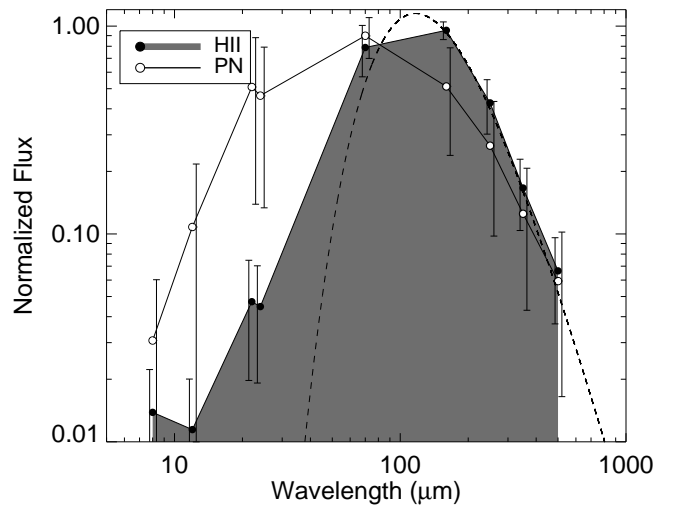}}

\caption{Spectral Energy Distributions.  Shown are the average
  normalized spectral energy distributions for \hiismall\ regions
  (dark grey filled area) and PNe (black curve).  The SED for dust of
  temperature 24.8\,K is overplotted with a dashed line.   Error
    bars show the standard deviation in each data point, and are to
    the left of the points for \hiismall\ regions and to the right for
    PNe.  There is considerable difference between the two average
  SEDs of \hiismall\ regions and PNe.}

\label{fig:sed}
\end{figure}

The average SEDs normalized to the peak flux for PNe and \hii\ regions
are shown in Figure~\ref{fig:sed}.  We only include the 12\,\micron,
22\,\micron, and 24\,\micron\ fluxes in the averages if the sources
are not saturated at these wavelengths according to our saturation
criterion.  For PNe not detected at all wavelengths, we exclude from
the averages the contribution from wavelengths where the source was
not detected.  The IR emission from PNe evolves with time as the dust
in the circumstellar shell expands \citep[e.g][]{garcia-lario03}.
Even for the same evolutionary state, whether a given PN is
oxygen-rich or carbon-rich can affect its IR properties \citep{volk03,
  bernard-salas06}.  Thus there is significantly more dispersion in
the average PN SED compared to that of the \hii\ regions,  as can
  be seen in the larger error bars in Figure~\ref{fig:sed}.

The difference between the average \hii\ region and PNe SEDs is clear.
For \hii\ regions, the SED peak near 100\,\micron\ can be ascribed to
``cold'' dust, while the MIR emission has contributions from
warmer dust and emission lines (largely PAHs).  The average PN SED
rises steeply from 8.0\,\micron\ to 24\,\micron, has a broad peak near
70\,\micron, and decreases from 70\,\micron\ long-ward.  We find that
a single temperature cannot fit the broad average SED of PNe,
indicating that there is dust at a range of temperatures.   The
  average SED for PNe looks similar to what is expected based on the
  SEDs of individual PNe in \citet{hora08} from 8.0\,\micron\ to
  70\,\micron.  There is, however, considerably less scatter in our
  70\,\micron\ data point than would be expected from their data.
  Additionally, many of their sources have higher 24\,\micron\ fluxes
  than 70\,\micron\ fluxes.

The 12\,\micron\ WISE flux for \hii\ regions is on average less than
the 8.0\,\micron\ GLIMPSE flux.  The average ratio $F_{12} / F_{8}$ is
$0.84\pm 0.26$.  We do not believe this is a systematic error in the
WISE 12\,\micron\ fluxes because no such trend was found when
comparing WISE fluxes with IRAS fluxes (see
Figure~\ref{fig:iras_wise}), but rather is caused by the PAH emission
in the 8.0\,\micron\ band (see also \S\ref{sec:colors}).

We find that the ``cold'' component of the mean \hii\ region SED is
well-fit by a single temperature grey body model with a dust
temperature $24.8\pm0.2$\,K, when the spectral index of the dust
emissivity, $\beta$, is held fixed to a value of 2.0.  In
these fits, we have assumed 25\% of the emission at 70\,\micron\ is
due to a hotter dust component.  Relaxing the assumption at
70\,\micron\ changes the derived temperatures to 26.3\,K, although the
fit quality is significantly worse.  The mean \hii\ region SED shown
in Figure~\ref{fig:sed} is very similar to the average SED for
``bubble'' \hii\ regions shown in \citet{anderson11b}, as are the
derived dust temperature values.

Over two-thirds (68\%) of \hii\ regions in our sample peak at
160\,\micron, while one third (32\%) peak at 70\,\micron.  Of the PNe
SEDs, 67\% peak at 70\,\micron, 23\% peak at either 22\,\micron\ or
24\,\micron, and 9\% (two sources) peak at 160\,\micron.  The two PNe
that peak at 160\,\micron\ are PNG002.2+00.5 and PNG358.6+00.7.

\begin{figure*}
\includegraphics[width=2.4 in]{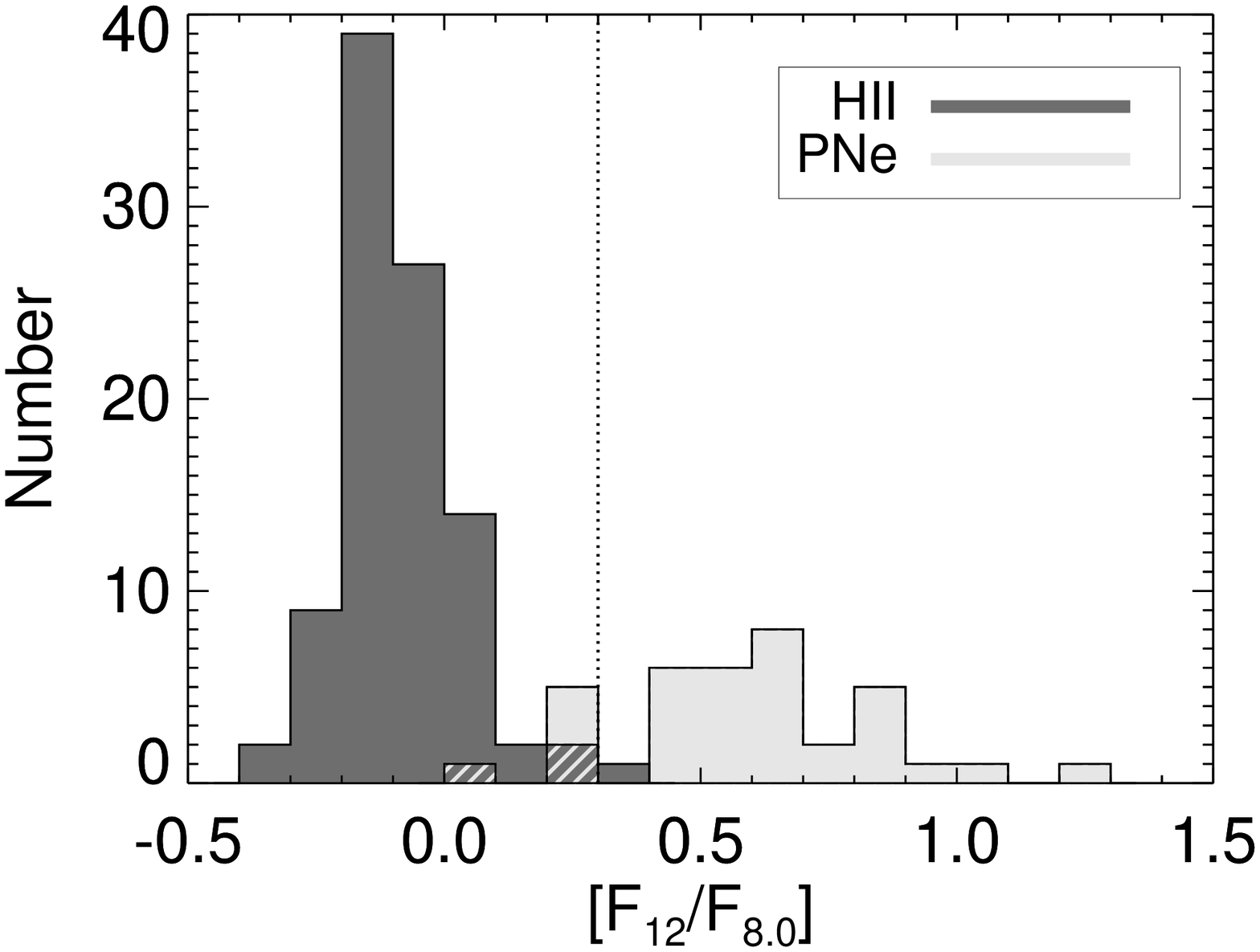}
\includegraphics[width=2.4 in]{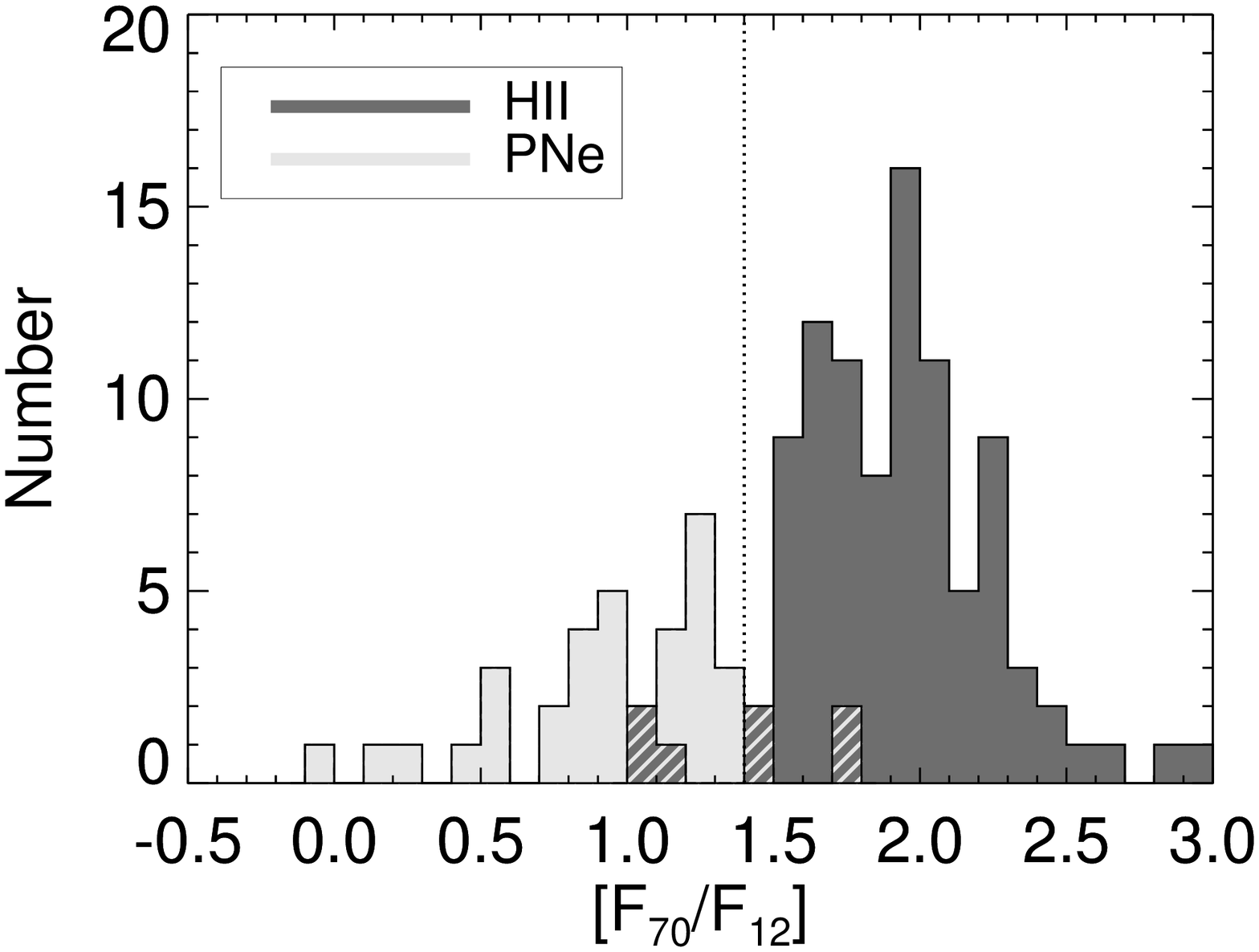}
\includegraphics[width=2.4 in]{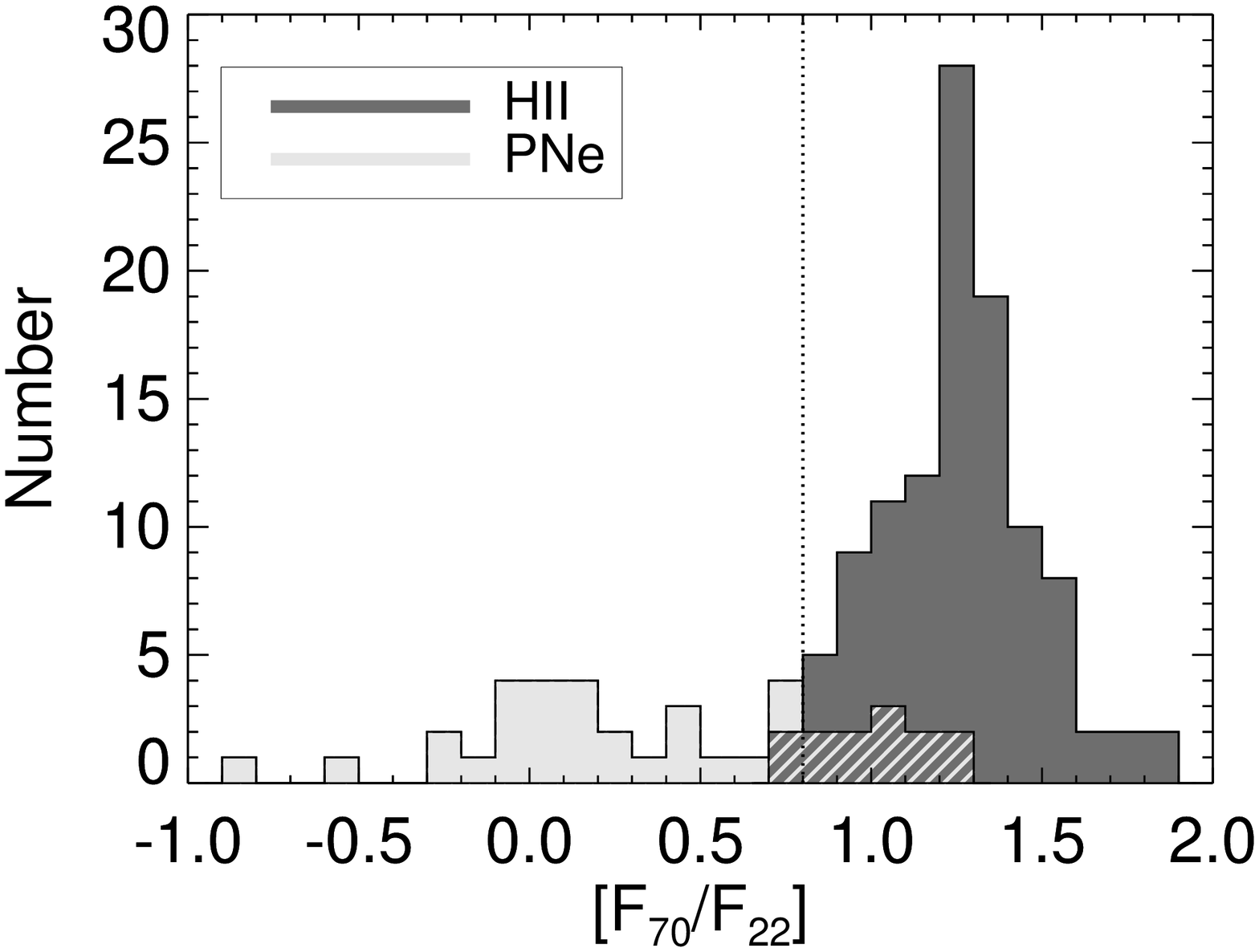}
\includegraphics[width=2.4 in]{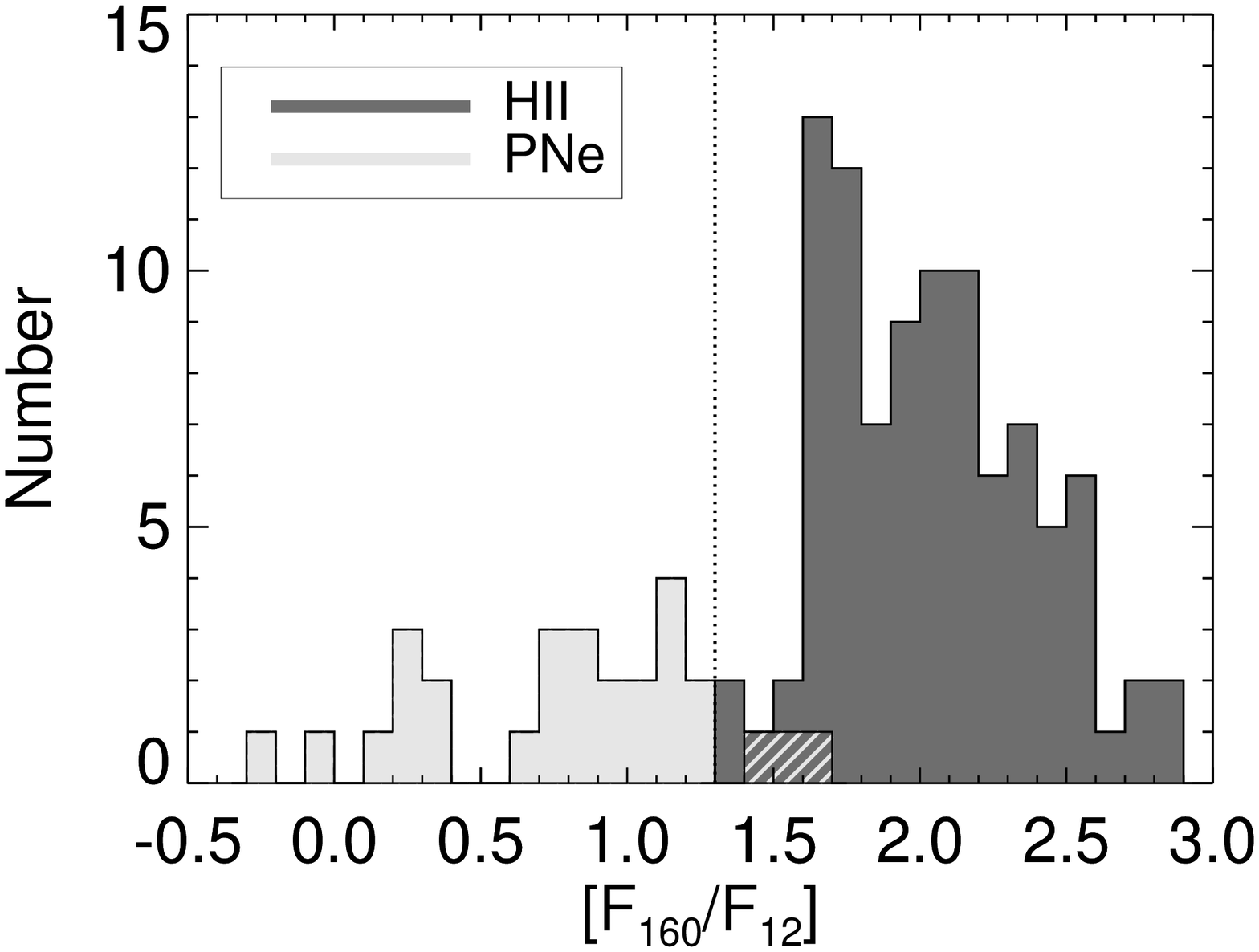}
\includegraphics[width=2.4 in]{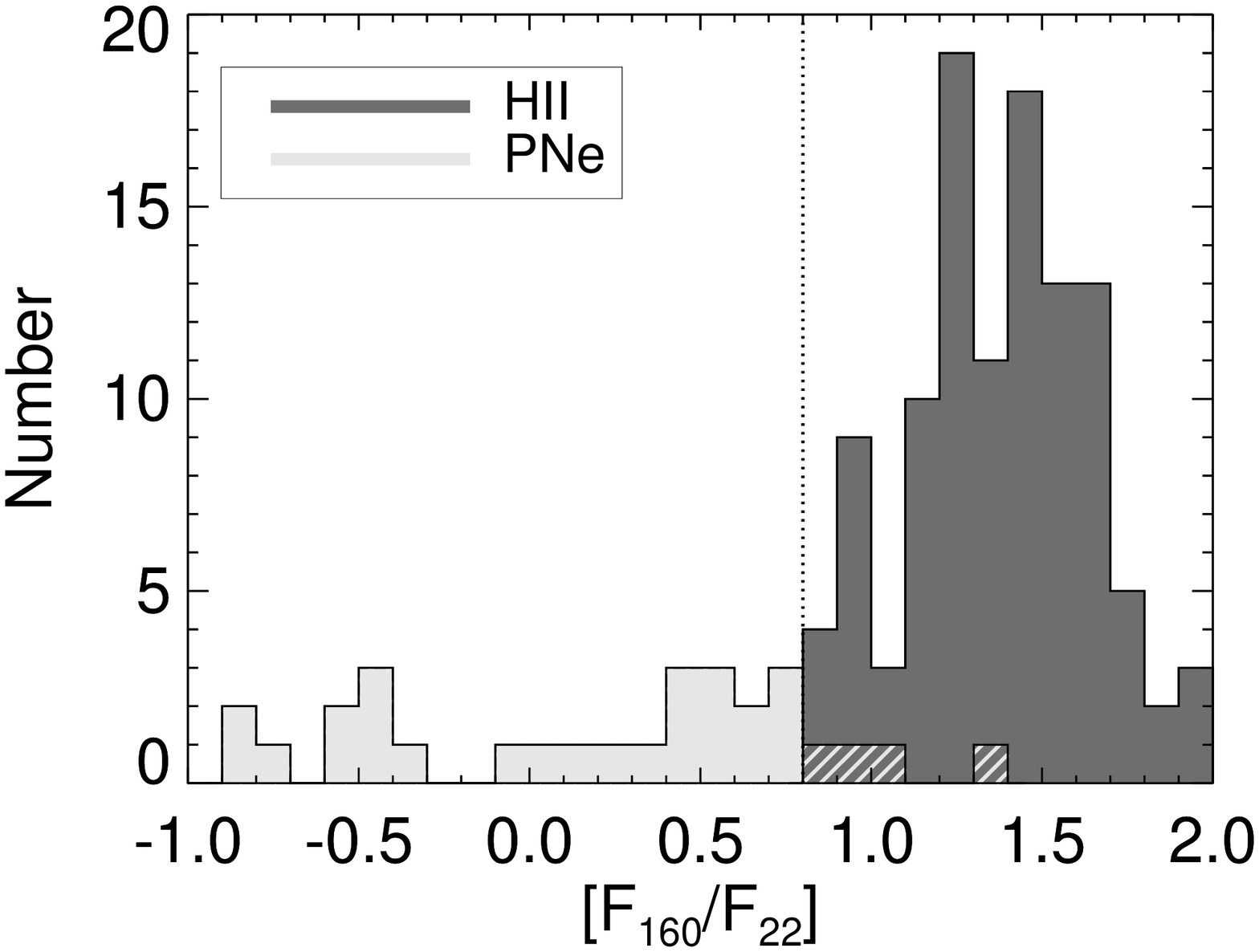}
\includegraphics[width=2.4 in]{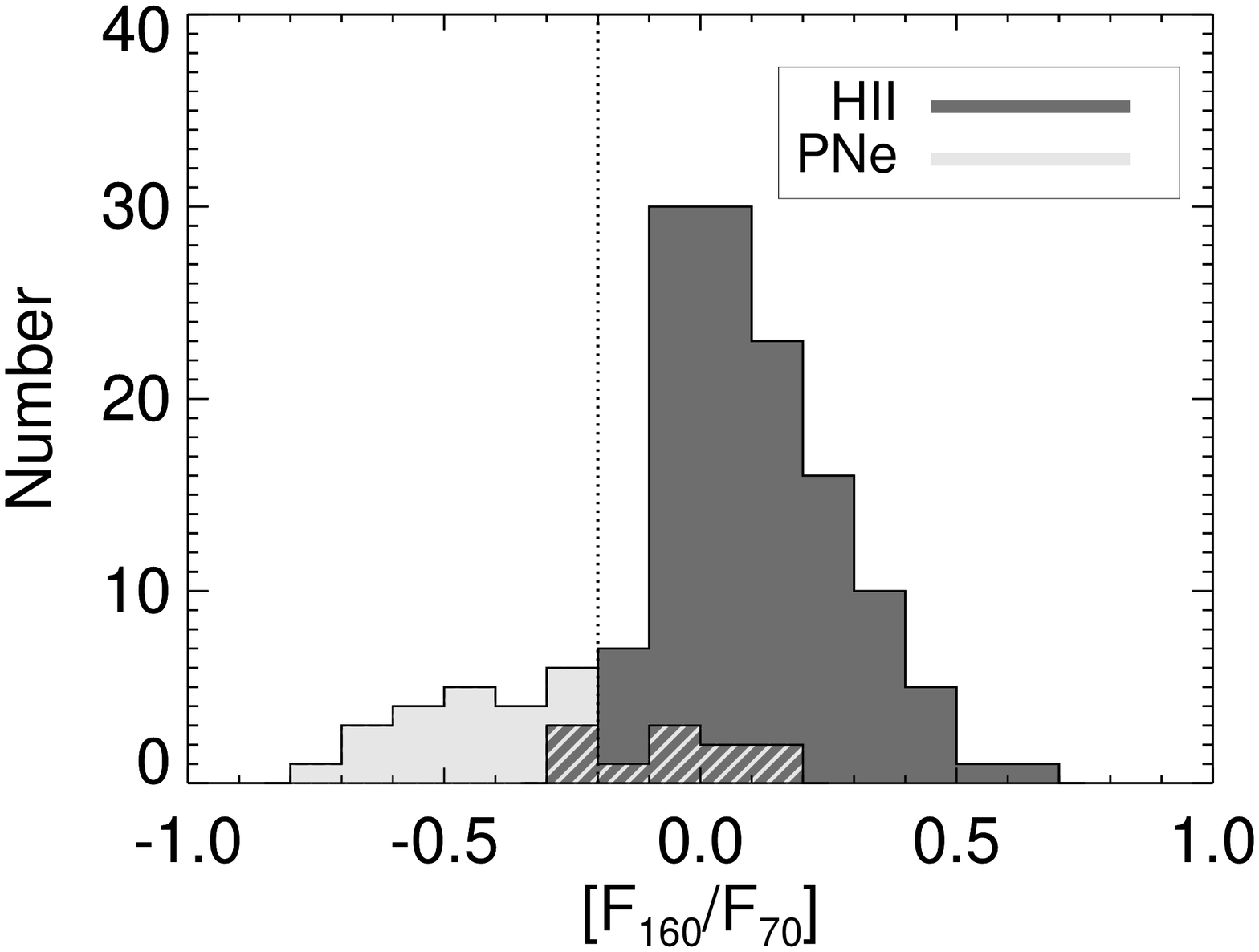}

\caption{IR color distributions for \hiismall\ regions and PNe.
  Shown are the six colors (the log of the flux ratio) best able to
  discriminate between \hiismall\ regions (dark grey) and PNe (light
  grey).  The hatched areas contain contributions from both
  \hiismall\ regions and PNe.  The vertical dotted lines show our
  proposed cutoff value for separating the two populations.}
\label{fig:ratios}
\end{figure*}

{\tiny
\begin{table*}
\caption{IR colors for disinguishing \hiismall\ regions from PNe.}
\label{tab:ratios}
\centering
\begin{tabular}{lrrrrrrr}
\hline\hline \\ [-2ex]
Color & Cutoff & \%H{\footnotesize\,II} & \%PNe &
Avg. (H{\footnotesize\,II}) & Avg. (PNe) & \#H{\footnotesize\,II}\ & \#PNe \\
\hline
$[F_{12}/F_{8}]$ & $< 0.3$ &  98 &  18 & $-0.09\pm0.11$ & $0.57\pm0.29$ &  96 &  38 \\
$[F_{22}/F_{8}]$ & $< 1.0$ &  92 &  20 & $0.59\pm0.31$ & $1.23\pm0.34$ & 110 &  39 \\
$[F_{24}/F_{8}]$ & $< 1.0$ &  93 &  21 & $0.55\pm0.29$ & $1.23\pm0.31$ &  86 &  37 \\
$[F_{70}/F_{12}]$ & $> 1.3$ &  96 &  20 & $1.92\pm0.36$ & $1.01\pm0.40$ &  96 &  40 \\
$[F_{70}/F_{22}]$ & $> 0.8$ &  98 &  26 & $1.25\pm0.22$ & $0.39\pm0.51$ & 110 &  42 \\
$[F_{70}/F_{24}]$ & $> 0.8$ & 100 &  26 & $1.26\pm0.21$ & $0.42\pm0.47$ &  86 &  41 \\
$[F_{160}/F_{12}]$ & $> 1.3$ & 100 &  10 & $2.06\pm0.38$ & $0.80\pm0.49$ &  96 &  30 \\
$[F_{160}/F_{22}]$ & $> 0.8$ & 100 &  12 & $1.38\pm0.26$ & $0.16\pm0.64$ & 110 &  31 \\
$[F_{160}/F_{24}]$ & $> 0.8$ & 100 &  10 & $1.40\pm0.25$ & $0.22\pm0.55$ &  86 &  29 \\
$[F_{160}/F_{70}]$ & $>-0.2$ &  97 &  25 & $0.10\pm0.17$ & $-0.32\pm0.24$ & 126 &  31 \\

\hline
\end{tabular}
\end{table*}
}

\subsection{Discriminating Colors\label{sec:colors}}
We examined the distributions of all possible colors to search for the
most robust criteria for discriminating between \hii\ regions and PNe.
The distributions for the most robust colors are shown in
Figure~\ref{fig:ratios}.  There, the hatched
areas contain contributions from both \hii\ regions and PNe.  The
vertical dotted lines in Figure~\ref{fig:ratios} are our proposed
discriminating color values.  Throughout, we use the notation such
that flux ratios enclosed in square brackets represent the log of the
flux ratio.

In Table~\ref{tab:ratios}, we list statistics on the colors shown in
Figure~\ref{fig:ratios}, as well the statistics of some additional
colors.  In Figure~\ref{fig:ratios}, we show colors using the WISE
22\,\micron\ data in lieu of the MIPSGAL 24\,\micron\ data, but we
list statistics on colors found using both data sets in
Table~\ref{tab:ratios}.  Table~\ref{tab:ratios} gives the color, the
proposed discriminating color cutoff value, the percentage of
\hii\ regions and PNe satisfying this color cutoff, the average value
of this color for the \hii\ regions and PNe in our sample, together
with their standard deviations, and the number of \hii\ regions and
PNe that were used in the calculations.  Table~\ref{tab:ratios} shows
that there are many different colors that can robustly separate PNe
and \hii\ regions.  The three most effective colors are
$[F_{12}/F_{8}]<0.3$, which is satisfied by 98\% of \hii\ regions and
7\% of PNe, $[F_{160}/F_{12}]>1.3$, which is satisfied by 100\% of
\hii\ regions and 10\% of PNe, and $[F_{160}/F_{24}]>0.8$, which is
also satisfied by 100\% of \hii\ regions and 10\% of PNe
($[F_{160}/F_{22}]>0.8$ is satisfied by 100\% of \hii\ regions and
12\% of PNe).

Table~\ref{tab:ratios} shows that \hii\ regions possess a relatively
narrow range of colors.  Since our sample has \hii\ regions in a
range of evolutionary states, any evolutionary effects in these colors
must be relatively modest.  The colors of PNe span a broader range of
values compared to those of \hii\ regions, for all colors shown in
Table~\ref{tab:ratios}.

Most of the variation in PNe colors is real and not the result of
photometric uncertainties.  For example, the standard deviation in
$[F_{160}/F_{24}]$ is 0.55 dex for PNe.  Restricting the calculation to
the 20 PNe with the highest 160\,\micron\ flux for which the
photometric uncertainty is relatively low, the standard deviation is
0.50; it is 0.52 for the 10 brightest PNe.  Similarly, the
$[F_{70}/F_{24}]$ color has a standard deviation of 0.47 for PNe; this
standard deviation is 0.43 and 0.38 for the 20 and 10 brightest PNe at
70\,\micron, respectively.

The colors that we have found most useful for discriminating between
\hii\ regions and PNe can be divided into three categories.  First,
many useful colors compare the flux at 70\,\micron\ or
160\,\micron\ with the flux in the MIR at 12\,\micron, 22\,\micron, or
24\,\micron.  A simple interpretation of these colors is that they
compare the emission from cold dust to that of a warmer dust
component.  Figure~\ref{fig:sed} shows that the average PN SED has
significantly more emission in the MIR.  This suggests that,
relative to the cold dust component, PNe have more emission from warm
dust than do \hii\ regions.  The second category are the colors
comparing 12\,\micron, 22\,\micron, or 24\,\micron\ fluxes with
8\,\micron\ fluxes.  This ratio is sensitive to the emission from
PAHs, which emit strongly at 8.0\,\micron, and less strongly in the
other bands.  As stated earlier, PAHs emit much more strongly for
\hii\ regions than for PNe.  Finally, the ratio $[F_{160}/F_{70}]$ is
  sensitive to the location of the SED peak.  Since on average the SEDs of
  \hii\ regions peak at longer wavelengths than PNe, this ratio will
  be positive for most \hii\ regions and negative for most PNe.


\begin{figure*}
\includegraphics[width=3.6 in]{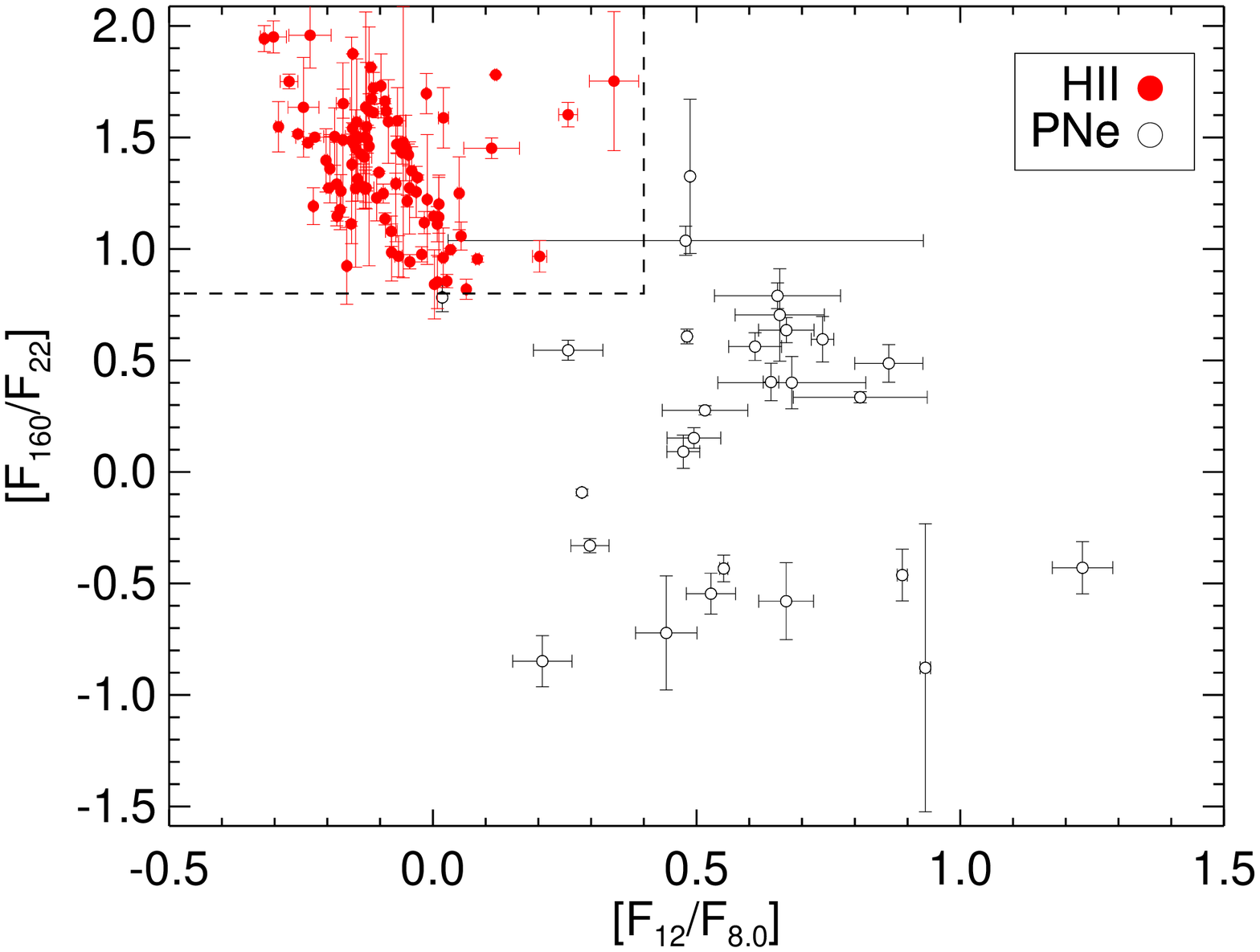}
\includegraphics[width=3.6 in]{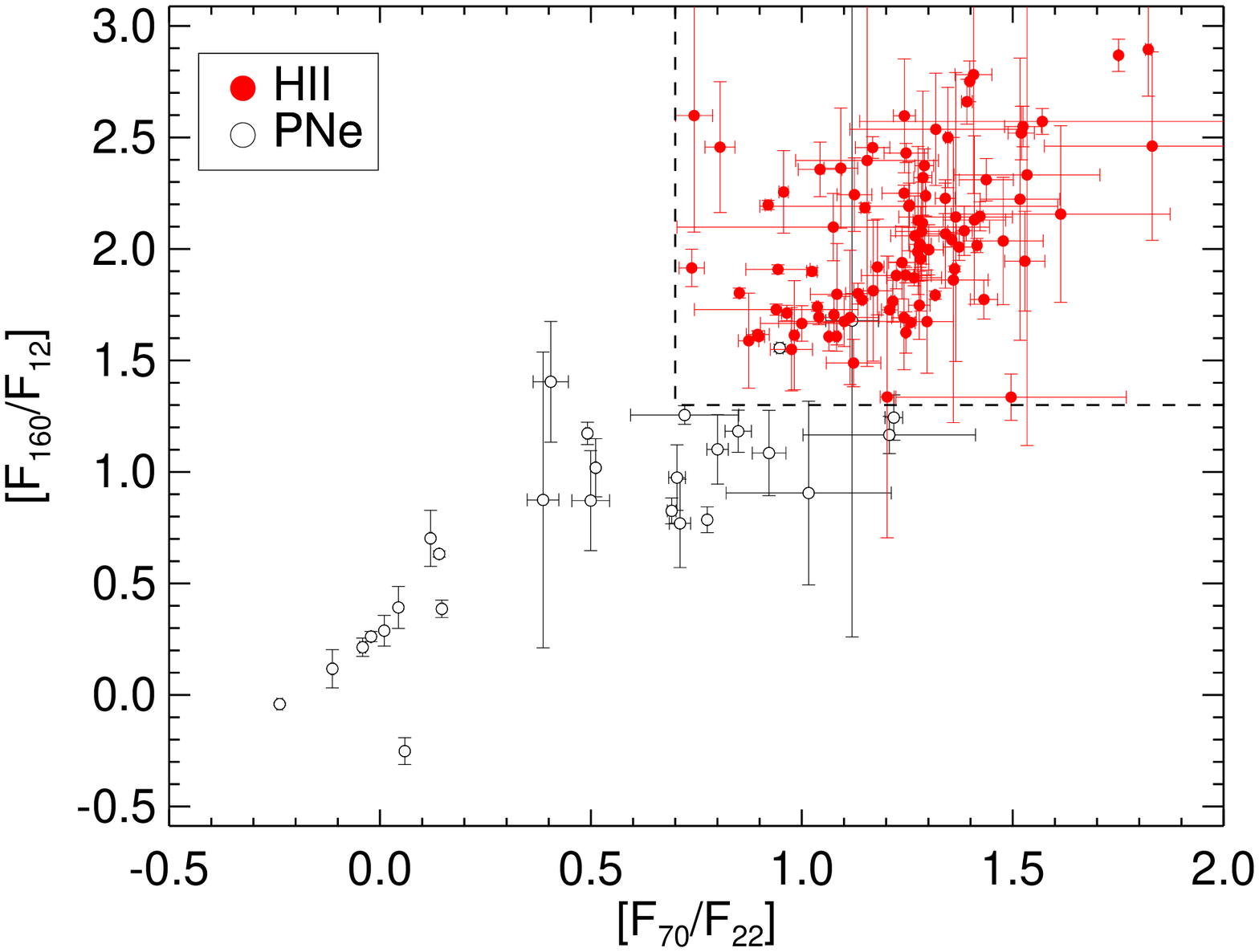}

\caption{Color-color diagrams for \hiismall\ regions and PNe.  The
  left panel shows the colors $[F_{12}/F_{8}]$ versus
  $[F_{160}/F_{22}]$ while the right panel shows $[F_{70}/F_{22}]$
  versus $[F_{160}/F_{12}]$.  The dashed lines in both panels enclose
  all \hiismall\ regions and few PNe: none in the left panel and two
  in the right panel.}
\label{fig:color}
\end{figure*}

\subsection{Color-color Diagrams}
There are numerous color combinations where \hii\ regions and PNe are
strongly segregated; the utility of an individual color-color plot
depends on the available measurements and their accuracy.  We show in
Figure~\ref{fig:color} color-color diagrams for two of the most robust
discriminating colors.  The left panel of Figure~\ref{fig:color} shows
the colors $[F_{12}/F_{8}]$ versus $[F_{160}/F_{22}]$ while the right
panel shows $[F_{70}/F_{22}]$ versus $[F_{160}/F_{12}]$.  The black
dashed lines in each panel enclose 100\% of our sample \hii\ regions.  In the
left panel, the area enclosed by these dashed lines includes no PNe
while in the right panel, only two PNe are within the area.  The
combination of two colors therefore shows a modest improvement in
discriminating power compared to the single colors alone.

 \section{Application to Objects of Unknown Classification} There
  are numerous astrophysical objects that the criteria here derived
  can hopefully help classify.  The power of these criteria is of
  course limited to separating \hii\ regions from PNe, and may not be
  applicable if the object in question is neither an \hii\ region nor
  a PN.  Below, we test our derived criteria on additional samples of
  objects of unknown or contested classification.  We also test the
  criteria on a sample of known PNe that are outside the coverage of
  Hi-GAL.

  Some of the objects listed as PNe in \citet{kwok08} are not included
  in other catalogs of PNe.  Most such objects appear visually similar
  to \hii\ regions: they have diffuse MIR emission extended over
  multiple arcminutes and show poor symmetry.  We perform aperture
  photometry from 8.0\,\micron\ to 500\,\micron\ as before (see
  \S\ref{sec:apphot}) on seven such nebulae that fall within the
  Galactic zone studied here.  These seven nebulae are listed in
  \citet{kwok08} as PNG298.4+00.6, PNG301.2+00.4, PNG321.0$-$00.7,
  PNG329.6$-$00.4, PNG332.5$-$00.1, PNG333.7+00.3, and PNG340.0+00.9.
  We find that six of these sources satisfy all the criteria listed in
  Table~\ref{tab:ratios} for \hii\ regions; their IR colors support
  their assignment as \hii\ regions rather than PNe.  PNG321.0$-$00.7,
  however, has colors more similar to PNe. For example, its ratio
  $[F_{160}/F_{12}]$ is 0.65.  This source is only a few arcseconds in
  diameter and seems likely to be a PN.

  Many of the ``disk and ring'' sources seen in MIPSGAL that were
  cataloged by \citet{mizuno10} do not have definitive
  classifications.  \citet{mizuno10} make a strong case that the
  catalog consists primarily of evolved stars, and therefore we would
  expect few \hii\ regions.  They list in their Appendix, however, two
  objects that based on their morphologies may be previously
  unclassified spiral galaxies.  One of these, MGE351.2381$-$00.0145,
  shown in their Figure~12, appears visually similar to an
  \hii\ region.  For this object, the 24\,\micron\ emission is
  surrounded by an 8.0\,\micron\ ring.  The morphology of
  MGE351.2381$-$00.0145 is similar to the \hii\ region G034.325+0.211
  detected in the HRDS\footnote{{\tiny see
    http://www.cv.nrao.edu/hrds/three\_color/three\_color\_LA229.jpg}}.
  The IR colors of MGE351.2381$-$00.0145 are consistant with it being an
  \hii\ region: its ratio $[F_{12}/F_{8}]$ is $-0.42$, and
  $[F_{22}/F_{8}]$ is $-0.39$.  We propose that this is an
  \hii\ region, and the ``spiral'' geometry arises from line-of-sight
  effects, as it likely also does for G034.325+0.211.  An additional
  object, named MGE031.7290+00.6993 in \citet{mizuno10}, is also
  included in the HRDS as G031.727+0.698.  This object also has
  24\,\micron\ emission surrounded by
  8.0\,\micron\ emission\footnote{{\tiny see
    http://www.cv.nrao.edu/hrds/three\_color/three\_color\_LA014.jpg}}.
  Its ratio $[F_{12}/F_{8}]$ is $-0.58$, and $[F_{22}/F_{8}]$ is
  $-0.13$, consistant with it being an \hii\ region.  Neither object
  is included in the Hi-GAL data range here studied\footnote{Between
    submission and publication we analyzed additional Hi-GAL data that
    were not originally available.  Both sources satisfy all the
    criteria for \hiismall\ regions in Table~\ref{tab:ratios}.}.

  There are 14 HRDS sources that appear point-like at 8.0\,\micron.
  Based on their morphology, they are likely PNe rather than
  \hii\ regions, although the line widths of their recombination line
  emission suggest that what was detected does not arise from PNe
  \citep[see][]{anderson11a}.  None of these sources lie within the
  Hi-GAL range studied here.  Of these 14 sources, only three satisfy
  $[F_{12}/F_{8}] > 0.3$, and eight satisfy $[F_{22}/F_{8}] > 1.0$,
  the colors associated with PNe.  Based on these colors, some of
  these sources many in fact be \hii\ regions.  Without Hi-GAL data,
  however, our discriminating power is more limited (see
  Table~\ref{tab:ratios}).

  Finally, there are a number of PNe near the Galactic Center region
  that have {\it Spitzer} data but, because of their Galactic
  latitudes, do not have Hi-GAL data.  We analyze 11 such objects,
  using only GLIMPSE, MIPSGAL, and WISE data.  We find that of these
  11, seven satisfy $[F_{12}/F_{8}] > 0.3$ and $[F_{22}/F_{8}] > 1.0$,
  consistant with the colors derived for PNe.

  The low success rates above show the importance of Hi-GAL data for
  such studies.  The central white dwarf of PNe can be detected at MIR
  wavelengths.  The contribution from the central source, relative to
  any contribution from a dusty shell, increases with decreasing
  wavelength.  We hypothesize that the flux from the central source
  decreases the discriminatory power of IR colors at lower
  wavelengths.

\section{Summary}
Using data from {\it Herschel} Hi-GAL, WISE, {\it Spitzer} MIPSGAL,
and {\it Spitzer} GLIMPSE, we have searched for the most robust
colors for separating a sample of 126 \hii\ regions from a sample of
43 planetary nebulae (PNe).   While our sample of PNe, which is
  confined to the Galactic plane and infrared bright, may not be
  representative of the entire Galactic population, it should be
  representative of PNe that may be confused with \hii\ regions.

\hii\ regions and PNe detected at MIR and FIR wavelengths can be
reliably separated using these data, and numerous colors can be used
to do so.  The best color criteria are $[F_{12}/F_{8}] < 0.3$, which
is satisfied by 98\% of our sample of \hii\ regions and 7\% of our
sample of PNe, $[F_{160}/F_{12}] > 1.3$, which is satisfied by 100\%
of our sample of \hii\ regions and 10\% of our sample of PNe, and
$[F_{160}/F_{24}] > 0.8$ which is also satisfied by 100\% of our
sample of \hii\ regions and 10\% of our sample of PNe.   Long
  wavelength Hi-GAL data greatly increase the discrimatory power of
  such IR color criteria.  We applied these criteria to objects of
  unknown classification and were able to discriminate between
  \hii\ regions and PNe in most cases.  Replacing the
MIPSGAL~24\,\micron\ fluxes with those of the WISE~22\,\micron\ band
in the previous ratio produces a nearly identical result.  The colors
of \hii\ regions span a relatively narrow range of values.  Any
evolution in these \hii\ region colors with age must therefore be
relatively small.

Combinations of these colors can be more effective at separating
\hii\ regions and PNe.  For example the combination $[F_{12}/F_{8}] <
0.4$ and $[F_{160}/F_{22}]>0.8$ includes all \hii\ regions in our
sample and no PNe.

The average SEDs of the \hii\ regions and PNe in our sample are
significantly different.  The average SED of \hii\ regions has an
obvious ``cold'' component that peaks near 100\,\micron, as well as
warmer component(s) that emit in the MIR and a strong contribution at
8.0\,\micron\ that may be caused by PAHs.  The average SED of PNe is
much broader than that of \hii\ regions and from these data cannot be
separated cleanly into temperature components.  The ``cold'' component
of the \hii\ region SED is well-fit with a grey-body of dust
temperature 25\,K.  The SED of over two-thirds of \hii\ regions peaks
at 160\,\micron, while the remaining one third peak at 70\,\micron.
The SEDs of 67\% of PNe peak at 70\,\micron, with the remainder
peaking at either 22\,\micron\ or 24\,\micron\ (23\%) or
160\,\micron\ (9\%).

We find that WISE~22\,\micron\ fluxes are strongly correlated with
MIPSGAL~24\,\micron\ fluxes over five orders of magnitude.  This
suggests that the WISE preliminary release data have sufficient
accuracy for the photometric measurements carried out here.  The
WISE~12\,\micron\ data are also well-correlated with the
IRAS~12\,\micron\ data, although the scatter is larger in part due to
the larger uncertainties when performing aperture photometry with IRAS
data.

These results are important for future work on \hii\ regions and PNe
since the two populations can be reliably separated using photometric
measurements and the discriminating IR colors derived here.  These
criteria will be useful when defining samples of \hii\ regions and PNe
from large-angle sky surveys.


\vskip 20pt We would like to thank the Hi-GAL team for their
continuing work on the survey.  PACS has been developed by a
consortium of institutes led by MPE (Germany) and including UVIE
(Austria); KU Leuven, CSL, IMEC (Belgium); CEA, LAM (France); MPIA
(Germany); INAF-IFSI/OAA/OAP/OAT, LENS, SISSA (Italy); IAC
(Spain). This development has been supported by the funding agencies
BMVIT (Austria), ESA-PRODEX (Belgium), CEA/CNES (France), DLR
(Germany), ASI/INAF (Italy), and CICYT/MCYT (Spain).  SPIRE has been
developed by a consortium of institutes led by Cardiff University (UK)
and including Univ. Lethbridge (Canada); NAOC (China); CEA, LAM
(France); IFSI, Univ. Padua (Italy); IAC (Spain); Stockholm
Observatory (Sweden); Imperial College London, RAL, UCL-MSSL, UKATC,
Univ. Sussex (UK); and Caltech, JPL, NHSC, Univ. Colorado (USA). This
development has been supported by national funding agencies: CSA
(Canada); NAOC (China); CEA, CNES, CNRS (France); ASI (Italy); MCINN
(Spain); SNSB (Sweden); STFC (UK); and NASA (USA). This publication
makes use of data products from the Wide-field Infrared Survey
Explorer, which is a joint project of the University of California,
Los Angeles, and the Jet Propulsion Laboratory/California Institute of
Technology, funded by the National Aeronautics and Space
Administration.  This research has made use of NASA's Astrophysics
Data System Bibliographic Services and the SIMBAD database operated at
CDS, Strasbourg, France.  L.D.A. acknowledges support by the ANR
Agence Nationale for the research project ``PROBeS'', number
ANR-08-BLAN-0241.

\bibliographystyle{aa} 
\bibliography{ref} 

\clearpage
\renewcommand{\tabcolsep}{2.5pt}
\longtab{1}{
\begin{landscape}
\begin{tiny}
\begin{longtable}{lrrrrrrrrrrrrrrrrrrrrrrl}
\caption{\label{tab:hii} The fluxes of \hiismall\ regions from 8.0\,\micron\ to 500\,\micron.}\\
\hline\hline \\ [-2ex]
Name & $\ell$\phantom{00} & $b$\phantom{00} & Size & $F_{8}$\,\, & $\sigma_{8}$ & $F_{12}$\,\, & $\sigma_{12}$ & $F_{22}$\,\, & $\sigma_{22}$ & $F_{24}$\,\, & $\sigma_{24}$ & $F_{70}$\,\, & $\sigma_{70}$ & $F_{160}$\, & $\sigma_{160}$ & $F_{250}$\, & $\sigma_{250}$ & $F_{350}$\, & $\sigma_{350}$ & $F_{500}$\, & $\sigma_{500}$ & Reference\\
 & deg.\,               & deg.\,            & $\arcmin$   & Jy\,\,     & Jy\,\,      & Jy\,\,      & Jy\,\,      & Jy\,\,      & Jy\,\,      & Jy\,\,      & Jy\,\,       & Jy\,\,       & Jy\,\,       & Jy\,\,       & Jy\,\,     & Jy\,\,      & Jy\,\,      & Jy\,\,       & Jy\,\,       & Jy\,\,       & Jy\,\, &           \\
\hline \\ [-2ex]
\endfirsthead
\caption{continued.}\\
\hline\hline \\ [-2ex]
Name & $\ell$\phantom{00} & $b$\phantom{00} & Size & $F_{8}$\,\, & $\sigma_{8}$ & $F_{12}$\,\, & $\sigma_{12}$ & $F_{22}$\,\, & $\sigma_{22}$ & $F_{24}$\,\, & $\sigma_{24}$ & $F_{70}$\,\, & $\sigma_{70}$ & $F_{160}$\, & $\sigma_{160}$ & $F_{250}$\, & $\sigma_{250}$ & $F_{350}$\, & $\sigma_{350}$ & $F_{500}$\, & $\sigma_{500}$ & Reference\\
 & deg.\,               & deg.\,            & $\arcmin$   & Jy\,\,     & Jy\,\,      & Jy\,\,      & Jy\,\,      & Jy\,\,      & Jy\,\,      & Jy\,\,      & Jy\,\,       & Jy\,\,       & Jy\,\,       & Jy\,\,       & Jy\,\,     & Jy\,\,      & Jy\,\,      & Jy\,\,       & Jy\,\,       & Jy\,\,       & Jy\,\, &           \\
\hline
\endhead
\hline
\endfoot
G002.961$-$0.053 & 2.961 & $-$0.053 & 1.7 & 4.2 & 0.6 & 2.1$\phantom{\dagger}$ & 0.5 & 6.9$\phantom{\dagger}$ & 0.6 & 8.3$\phantom{\dagger}$ & 1.8 & 460 & 100.0 & 610 & 29.0 & 600 & 150.0 & 300 & 71.0 & 100.0 & 22.0 & HRDS \\
G002.611+0.135 & 2.611 & +0.135 & 13.0 & 6.6 & 1.3 & 5.4$\phantom{\dagger}$ & 0.8 & 51.0$\phantom{\dagger}$ & 3.1 & 46.0$\phantom{\dagger}$ & 2.9 & 890 & 73.0 & 2100 & 110.0 & 830 & 130.0 & 350 & 66.0 & 110.0 & 23.0 & Known \\
G002.404+0.068 & 2.404 & +0.068 & 6.5 & 5.1 & 0.2 & 3.6$\phantom{\dagger}$ & 0.2 & 18.0$\phantom{\dagger}$ & 1.2 & 19.0$\phantom{\dagger}$ & 1.0 & 150 & 9.2 & 570 & 45.0 & 130 & 6.0 & 52 & 3.4 & 17.0 & 1.4 & HRDS \\
G002.009$-$0.680 & 2.009 & $-$0.680 & 1.3 & 2.1 & 0.3 & 1.7$\phantom{\dagger}$ & 0.2 & 7.2$\phantom{\dagger}$ & 0.8 & 7.7$\phantom{\dagger}$ & 0.5 & 130 & 50.0 & 270 & 61.0 & 130 & 39.0 & 60 & 14.0 & 22.0 & 4.0 & HRDS \\
G001.488$-$0.199 & 1.488 & $-$0.199 & 1.3 & 1.7 & 0.1 & 0.8$\phantom{\dagger}$ & 0.1 & 3.6$\phantom{\dagger}$ & 0.1 & 4.4$\phantom{\dagger}$ & 0.1 & 130 & 40.0 & 310 & 18.0 & 140 & 47.0 & 65 & 22.0 & 21.0 & 7.1 & HRDS \\
G001.330+0.150 & 1.330 & +0.150 & 3.2 & 24.0 & 1.4 & 17.0$\phantom{\dagger}$ & 0.7 & 26.0$\phantom{\dagger}$ & 0.3 & 30.0$\phantom{\dagger}$ & 0.3 & 220 & 49.0 & 900 & 27.0 & 190 & 66.0 & 84 & 35.0 & 27.0 & 13.0 & HRDS \\
G000.838+0.189 & 0.838 & +0.189 & 1.8 & 43.0 & 15.0 & 30.0$\phantom{\dagger}$ & 10.0 & 120.0$\phantom{\dagger}$ & 12.0 & 120.0$\phantom{\dagger}$ & 20.0 & 1500 & 640.0 & 3700 & 1500.0 & 1300 & 580.0 & 560 & 260.0 & 200.0 & 93.0 & HRDS \\
G000.640+0.623 & 0.640 & +0.623 & 3.6 & 170.0 & 32.0 & 370.0$\phantom{\dagger}$ & 120.0 & 140.0$\phantom{\dagger}$ & 10.0 & 170.0$\phantom{\dagger}$ & 21.0 & 4500 & 1100.0 & 8000 & 2000.0 & 3600 & 910.0 & 1500 & 370.0 & 460.0 & 110.0 & Known \\
G000.382+0.017 & 0.382 & +0.017 & 2.4 & 17.0 & 8.1 & 13.0$\dagger$ & 6.5 & 84.0$\phantom{\dagger}$ & 8.8 & 65.0$\dagger$ & 8.8 & 800 & 190.0 & 2400 & 940.0 & 660 & 140.0 & 260 & 53.0 & 65.0 & 16.0 & HRDS \\
G359.277$-$0.264 & 359.277 & $-$0.264 & 3.7 & 95.0 & 20.0 & 71.0$\dagger$ & 8.1 & 300.0$\phantom{\dagger}$ & 35.0 & 280.0$\dagger$ & 6.1 & 4300 & 1300.0 & 13000 & 3000.0 & 3200 & 880.0 & 1100 & 290.0 & 320.0 & 79.0 & Known \\
G358.720+0.011 & 358.720 & +0.011 & 7.1 & 11.0 & 0.8 & 9.6$\phantom{\dagger}$ & 0.4 & 84.0$\phantom{\dagger}$ & 9.8 & 58.0$\phantom{\dagger}$ & 11.0 & 1000 & 83.0 & 2200 & 330.0 & 960 & 71.0 & 400 & 27.0 & 120.0 & 7.2 & HRDS \\
G358.633+0.062 & 358.633 & +0.062 & 3.1 & 5.7 & 0.4 & 4.6$\phantom{\dagger}$ & 0.4 & 23.0$\phantom{\dagger}$ & 4.1 & 25.0$\dagger$ & 5.0 & 400 & 90.0 & 1200 & 120.0 & 350 & 89.0 & 140 & 36.0 & 43.0 & 13.0 & HRDS \\
G348.550$-$0.339 & 348.550 & $-$0.339 & 2.3 & 0.5 & 0.1 & 0.4$\phantom{\dagger}$ & 0.1 & 4.7$\phantom{\dagger}$ & 0.3 & 5.0$\phantom{\dagger}$ & 0.1 & 82 & 1.0 & 77 & 5.8 & 31 & 3.8 & 11 & 1.5 & 4.3 & 0.5 & RMS \\
G348.230$-$0.970 & 348.230 & $-$0.970 & 11.9 & 460.0 & 14.0 & 300.0$\dagger$ & 4.5 & 1200.0$\dagger$ & 7.3 & 870.0$\dagger$ & 14.0 & 15000 & 290.0 & 13000 & 980.0 & 4900 & 520.0 & 1700 & 200.0 & 610.0 & 79.0 & Known \\
G348.225+0.459 & 348.225 & +0.459 & 3.0 & 490.0 & 68.0 & 420.0$\phantom{\dagger}$ & 15.0 & 1100.0$\phantom{\dagger}$ & 14.0 & 1000.0$\phantom{\dagger}$ & 43.0 & 14000 & 1000.0 & 21000 & 4400.0 & 10000 & 2400.0 & 4200 & 880.0 & 1800.0 & 360.0 & Known \\
G348.148+0.255 & 348.148 & +0.255 & 1.4 & 8.9 & 5.0 & 7.6$\phantom{\dagger}$ & 2.5 & 24.0$\phantom{\dagger}$ & 3.1 & 26.0$\phantom{\dagger}$ & 4.4 & 460 & 82.0 & 910 & 200.0 & 470 & 93.0 & 180 & 35.0 & 72.0 & 14.0 & HRDS \\
G347.964$-$0.439 & 347.964 & $-$0.439 & 3.4 & 150.0 & 5.0 & 110.0$\phantom{\dagger}$ & 17.0 & 300.0$\phantom{\dagger}$ & 8.1 & 330.0$\phantom{\dagger}$ & 5.8 & 3600 & 92.0 & 5700 & 570.0 & 2900 & 320.0 & 1100 & 120.0 & 470.0 & 49.0 & Known \\
G347.918$-$0.761 & 347.918 & $-$0.761 & 3.5 & 3.1 & 0.1 & 3.2$\phantom{\dagger}$ & 0.1 & 18.0$\phantom{\dagger}$ & 0.4 & 15.0$\phantom{\dagger}$ & 0.5 & 180 & 2.8 & 130 & 21.0 & 51 & 17.0 & 19 & 7.6 & 8.3 & 3.3 & HRDS \\
G347.536+0.084 & 347.536 & +0.084 & 3.2 & 160.0 & 46.0 & 110.0$\phantom{\dagger}$ & 30.0 & 290.0$\phantom{\dagger}$ & 37.0 & 300.0$\phantom{\dagger}$ & 31.0 & 5100 & 530.0 & 5400 & 2200.0 & 2200 & 710.0 & 740 & 250.0 & 260.0 & 88.0 & HRDS \\
G347.386+0.266 & 347.386 & +0.266 & 10.0 & 2000.0 & 120.0 & 1300.0$\phantom{\dagger}$ & 62.0 & 4600.0$\phantom{\dagger}$ & 220.0 & 4400.0$\dagger$ & 120.0 & 69000 & 2400.0 & 100000 & 17000.0 & 46000 & 10000.0 & 16000 & 3900.0 & 6400.0 & 1600.0 & Known \\
G346.875+0.328 & 346.875 & +0.328 & 2.1 & 4.4 & 1.3 & 4.2$\phantom{\dagger}$ & 0.7 & 8.8$\phantom{\dagger}$ & 2.1 & 10.0$\phantom{\dagger}$ & 2.7 & 230 & 53.0 & 440 & 81.0 & 220 & 33.0 & 84 & 12.0 & 35.0 & 4.6 & HRDS \\
G346.539+0.097 & 346.539 & +0.097 & 1.9 & 11.0 & 0.9 & 10.0$\phantom{\dagger}$ & 0.7 & 43.0$\phantom{\dagger}$ & 0.8 & 41.0$\phantom{\dagger}$ & 0.9 & 1000 & 33.0 & 1200 & 150.0 & 500 & 12.0 & 180 & 4.1 & 65.0 & 2.3 & Known \\
G346.233$-$0.320 & 346.233 & $-$0.320 & 18.5 & 4.3 & 0.8 & 3.0$\phantom{\dagger}$ & 0.6 & 13.0$\phantom{\dagger}$ & 0.4 & 12.0$\phantom{\dagger}$ & 0.8 & 250 & 15.0 & 300 & 46.0 & 140 & 25.0 & 54 & 9.4 & 21.0 & 3.6 & RMS \\
G346.099$-$0.364 & 346.099 & $-$0.364 & 2.5 & 13.0 & 3.2 & 10.0$\phantom{\dagger}$ & 1.9 & 13.0$\phantom{\dagger}$ & 1.9 & 14.0$\phantom{\dagger}$ & 3.0 & 260 & 50.0 & 470 & 160.0 & 210 & 83.0 & 69 & 15.0 & 30.0 & 13.0 & HRDS \\
G346.077$-$0.056 & 346.077 & $-$0.056 & 2.3 & 9.2 & 1.2 & 6.8$\phantom{\dagger}$ & 1.4 & 29.0$\phantom{\dagger}$ & 2.0 & 25.0$\phantom{\dagger}$ & 2.2 & 540 & 43.0 & 920 & 43.0 & 490 & 17.0 & 210 & 7.2 & 78.0 & 4.5 & HRDS \\
G345.722+0.153 & 345.722 & +0.153 & 4.1 & 22.0 & 1.0 & 15.0$\phantom{\dagger}$ & 0.5 & 120.0$\phantom{\dagger}$ & 2.7 & 130.0$\phantom{\dagger}$ & 3.3 & 1700 & 110.0 & 990 & 210.0 & 160 & 11.0 & 45 & 4.2 & 14.0 & 1.3 & HRDS \\
G345.490+0.310 & 345.490 & +0.310 & 3.7 & 69.0 & 4.9 & 43.0$\dagger$ & 3.2 & 210.0$\dagger$ & 5.6 & 180.0$\dagger$ & 4.4 & 9500 & 110.0 & 11000 & 160.0 & 4800 & 130.0 & 1700 & 31.0 & 790.0 & 29.0 & Known \\
G345.425$-$0.940 & 345.425 & $-$0.940 & 1.4 & 1100.0 & 30.0 & 470.0$\dagger$ & 8.0 & 2100.0$\dagger$ & 11.0 & 1700.0$\dagger$ & 11.0 & 51000 & 180.0 & 37000 & 1400.0 & 13000 & 880.0 & 3000 & 390.0 & 1800.0 & 150.0 & Known \\
G345.215$-$0.749 & 345.215 & $-$0.749 & 10.1 & 230.0 & 28.0 & 230.0$\phantom{\dagger}$ & 38.0 & 920.0$\phantom{\dagger}$ & 67.0 & 940.0$\phantom{\dagger}$ & 15.0 & 8400 & 510.0 & 12000 & 1700.0 & 5300 & 280.0 & 1100 & 150.0 & 920.0 & 40.0 & Known \\
G345.097+0.136 & 345.097 & +0.136 & 1.7 & 4.5 & 1.3 & 2.6$\phantom{\dagger}$ & 1.2 & 4.1$\phantom{\dagger}$ & 1.5 & 5.5$\phantom{\dagger}$ & 2.4 & 170 & 34.0 & 370 & 110.0 & 310 & 73.0 & 93 & 28.0 & 64.0 & 17.0 & HRDS \\
G345.004$-$0.224 & 345.004 & $-$0.224 & 2.9 & 2.4 & 0.1 & 1.7$\phantom{\dagger}$ & 0.1 & 73.0$\phantom{\dagger}$ & 0.5 & 31.0$\dagger$ & 0.5 & 4500 & 6.7 & 5500 & 38.0 & 2100 & 39.0 & 1100 & 21.0 & 400.0 & 9.2 & RMS \\
G344.439+0.048 & 344.439 & +0.048 & 2.8 & 61.0 & 1.2 & 35.0$\dagger$ & 1.2 & 190.0$\dagger$ & 1.5 & 160.0$\dagger$ & 2.4 & 5000 & 11.0 & 5100 & 83.0 & 2300 & 76.0 & 730 & 40.0 & 290.0 & 17.0 & Known \\
G344.226$-$0.588 & 344.226 & $-$0.588 & 2.4 & 35.0 & 0.8 & 19.0$\dagger$ & 0.7 & 110.0$\phantom{\dagger}$ & 1.0 & 73.0$\dagger$ & 0.9 & 3500 & 9.6 & 4800 & 69.0 & 2500 & 65.0 & 890 & 35.0 & 400.0 & 19.0 & Known \\
G344.106$-$0.672 & 344.106 & $-$0.672 & 1.2 & 26.0 & 2.4 & 15.0$\phantom{\dagger}$ & 3.1 & 58.0$\phantom{\dagger}$ & 2.5 & 62.0$\phantom{\dagger}$ & 1.6 & 1300 & 32.0 & 2500 & 450.0 & 1600 & 420.0 & 590 & 110.0 & 370.0 & 130.0 & HRDS \\
G343.910+0.115 & 343.910 & +0.115 & 7.3 & 7.3 & 3.8 & 5.4$\phantom{\dagger}$ & 2.7 & 17.0$\phantom{\dagger}$ & 1.0 & 18.0$\phantom{\dagger}$ & 1.9 & 440 & 49.0 & 730 & 240.0 & 380 & 200.0 & 170 & 97.0 & 82.0 & 42.0 & HRDS \\
G343.856$-$0.106 & 343.856 & $-$0.106 & 4.1 & 53.0 & 1.8 & 38.0$\phantom{\dagger}$ & 1.7 & 73.0$\phantom{\dagger}$ & 4.1 & 80.0$\phantom{\dagger}$ & 3.7 & 2000 & 68.0 & 2300 & 210.0 & 1100 & 84.0 & 470 & 15.0 & 200.0 & 5.4 & HRDS \\
G343.721$-$0.221 & 343.721 & $-$0.221 & 3.2 & 6.6 & 0.1 & 3.4$\phantom{\dagger}$ & 0.1 & 30.0$\phantom{\dagger}$ & 1.5 & 27.0$\phantom{\dagger}$ & 1.2 & 670 & 31.0 & 1100 & 130.0 & 590 & 86.0 & 250 & 37.0 & 100.0 & 16.0 & HRDS \\
G343.490$-$0.033 & 343.490 & $-$0.033 & 4.6 & 540.0 & 14.0 & 490.0$\dagger$ & 14.0 & 2300.0$\dagger$ & 56.0 & 2000.0$\dagger$ & 63.0 & 21000 & 610.0 & 18000 & 1700.0 & 6800 & 1100.0 & 2300 & 470.0 & 840.0 & 200.0 & Known \\
G342.061+0.420 & 342.061 & +0.420 & 2.7 & 73.0 & 20.0 & 71.0$\phantom{\dagger}$ & 10.0 & 270.0$\phantom{\dagger}$ & 7.9 & 270.0$\phantom{\dagger}$ & 11.0 & 3200 & 280.0 & 4500 & 1200.0 & 2200 & 670.0 & 880 & 250.0 & 380.0 & 100.0 & RMS \\
G340.777$-$1.008 & 340.777 & $-$1.008 & 2.2 & 250.0 & 20.0 & 130.0$\dagger$ & 9.4 & 560.0$\dagger$ & 20.0 & 450.0$\dagger$ & 13.0 & 14000 & 220.0 & 12000 & 300.0 & 5000 & 120.0 & 1700 & 49.0 & 620.0 & 23.0 & Known \\
G340.249$-$0.046 & 340.249 & $-$0.046 & 2.5 & 2.8 & 0.1 & 2.4$\phantom{\dagger}$ & 0.1 & 46.0$\phantom{\dagger}$ & 1.0 & 36.0$\phantom{\dagger}$ & 0.8 & 1100 & 21.0 & 1300 & 63.0 & 650 & 38.0 & 250 & 16.0 & 88.0 & 7.2 & RMS \\
G340.047$-$0.253 & 340.047 & $-$0.253 & 1.1 & 28.0 & 9.1 & 23.0$\dagger$ & 4.9 & 170.0$\phantom{\dagger}$ & 5.8 & 140.0$\dagger$ & 7.6 & 6400 & 230.0 & 6700 & 390.0 & 2600 & 170.0 & 1100 & 60.0 & 390.0 & 25.0 & Known \\
G339.760+0.053 & 339.760 & +0.053 & 1.8 & 5.5 & 0.4 & 3.9$\phantom{\dagger}$ & 0.3 & 8.2$\phantom{\dagger}$ & 0.1 & 9.4$\phantom{\dagger}$ & 0.1 & 170 & 2.8 & 240 & 19.0 & 110 & 12.0 & 39 & 5.4 & 15.0 & 2.4 & RMS \\
G339.747+0.095 & 339.747 & +0.095 & 3.2 & 1.2 & 0.2 & 0.6$\phantom{\dagger}$ & 0.1 & 6.7$\phantom{\dagger}$ & 0.5 & 7.5$\phantom{\dagger}$ & 0.5 & 170 & 6.1 & 380 & 18.0 & 220 & 15.0 & 88 & 7.6 & 36.0 & 3.9 & RMS \\
G339.104+0.148 & 339.104 & +0.148 & 7.2 & 10.0 & 1.3 & 10.0$\phantom{\dagger}$ & 0.7 & 60.0$\phantom{\dagger}$ & 0.9 & 45.0$\dagger$ & 0.9 & 640 & 17.0 & 790 & 47.0 & 390 & 36.0 & 150 & 16.0 & 55.0 & 6.3 & RMS \\
G338.681$-$0.084 & 338.681 & $-$0.084 & 5.2 & 1.6 & 0.1 & 1.4$\phantom{\dagger}$ & 0.1 & 16.0$\phantom{\dagger}$ & 0.8 & 12.0$\phantom{\dagger}$ & 0.6 & 300 & 20.0 & 300 & 62.0 & 120 & 29.0 & 40 & 10.0 & 14.0 & 3.7 & RMS \\
G338.405$-$0.203 & 338.405 & $-$0.203 & 1.9 & 7.2 & 1.1 & 11.0$\phantom{\dagger}$ & 1.0 & 66.0$\phantom{\dagger}$ & 1.6 & 49.0$\dagger$ & 1.6 & 1100 & 44.0 & 610 & 60.0 & 220 & 24.0 & 76 & 8.5 & 30.0 & 3.4 & RMS \\
G338.374$-$0.152 & 338.374 & $-$0.152 & 7.9 & 1.6 & 0.2 & 1.7$\phantom{\dagger}$ & 0.1 & 9.5$\phantom{\dagger}$ & 0.4 & 9.2$\phantom{\dagger}$ & 0.4 & 180 & 9.5 & 150 & 23.0 & 57 & 12.0 & 19 & 4.8 & 7.3 & 2.0 & RMS \\
G337.995+0.077 & 337.995 & +0.077 & 1.8 & 15.0 & 3.4 & 9.5$\phantom{\dagger}$ & 2.3 & 44.0$\phantom{\dagger}$ & 2.2 & 36.0$\phantom{\dagger}$ & 2.4 & 810 & 77.0 & 1100 & 170.0 & 570 & 90.0 & 220 & 34.0 & 89.0 & 13.0 & RMS \\
G337.949$-$0.476 & 337.949 & $-$0.476 & 6.1 & 490.0 & 65.0 & 350.0$\dagger$ & 7.9 & 1800.0$\dagger$ & 25.0 & 1300.0$\dagger$ & 49.0 & 35000 & 1200.0 & 34000 & 5200.0 & 14000 & 3000.0 & 5500 & 1200.0 & 2200.0 & 490.0 & Known \\
G337.004+0.324 & 337.004 & +0.324 & 3.9 & 3.3 & 0.2 & 2.4$\phantom{\dagger}$ & 0.3 & 22.0$\phantom{\dagger}$ & 0.7 & 20.0$\phantom{\dagger}$ & 2.7 & 460 & 85.0 & 820 & 170.0 & 420 & 68.0 & 160 & 23.0 & 62.0 & 8.6 & RMS \\
G336.773+0.246 & 336.773 & +0.246 & 2.2 & 7.6 & 2.7 & 6.3$\phantom{\dagger}$ & 1.4 & 16.0$\phantom{\dagger}$ & 1.4 & 18.0$\phantom{\dagger}$ & 1.5 & 210 & 31.0 & 190 & 29.0 & 75 & 9.8 & 25 & 3.3 & 8.7 & 1.3 & RMS \\
G335.197$-$0.389 & 335.197 & $-$0.389 & 3.7 & 1.4 & 0.3 & 0.9$\phantom{\dagger}$ & 0.3 & 4.6$\phantom{\dagger}$ & 0.4 & 6.0$\phantom{\dagger}$ & 0.5 & 150 & 3.9 & 140 & 14.0 & 60 & 8.5 & 23 & 3.8 & 8.7 & 1.9 & RMS \\
G334.714$-$0.665 & 334.714 & $-$0.665 & 1.8 & 12.0 & 0.3 & 14.0$\phantom{\dagger}$ & 0.9 & 63.0$\phantom{\dagger}$ & 0.8 & 54.0$\phantom{\dagger}$ & 1.6 & 730 & 7.2 & 570 & 16.0 & 220 & 12.0 & 80 & 6.5 & 32.0 & 3.6 & Known \\
G334.529+0.825 & 334.529 & +0.825 & 4.6 & 110.0 & 11.0 & 81.0$\phantom{\dagger}$ & 4.8 & 210.0$\phantom{\dagger}$ & 40.0 & 200.0$\phantom{\dagger}$ & 32.0 & 2600 & 140.0 & 3800 & 490.0 & 1500 & 270.0 & 520 & 100.0 & 200.0 & 43.0 & Known \\
G333.736$-$0.425 & 333.736 & $-$0.425 & 1.5 & 3.3 & 0.5 & 2.9$\phantom{\dagger}$ & 0.3 & 15.0$\phantom{\dagger}$ & 0.8 & 17.0$\phantom{\dagger}$ & 0.7 & 400 & 19.0 & 410 & 36.0 & 170 & 22.0 & 64 & 8.3 & 26.0 & 3.1 & RMS \\
G333.679$-$0.435 & 333.679 & $-$0.435 & 9.0 & 4.1 & 0.7 & 3.8$\phantom{\dagger}$ & 0.8 & 18.0$\phantom{\dagger}$ & 3.8 & 17.0$\phantom{\dagger}$ & 3.0 & 320 & 58.0 & 330 & 39.0 & 130 & 15.0 & 47 & 7.8 & 18.0 & 4.2 & RMS \\
G333.375$-$0.202 & 333.375 & $-$0.202 & 1.6 & 2.7 & 0.4 & 1.8$\phantom{\dagger}$ & 0.2 & 16.0$\phantom{\dagger}$ & 1.1 & 14.0$\phantom{\dagger}$ & 0.6 & 220 & 22.0 & 320 & 66.0 & 160 & 35.0 & 61 & 13.0 & 23.0 & 5.1 & RMS \\
G333.340$-$0.128 & 333.340 & $-$0.128 & 1.4 & 3.6 & 1.2 & 3.7$\phantom{\dagger}$ & 0.6 & 33.0$\phantom{\dagger}$ & 1.2 & 34.0$\phantom{\dagger}$ & 1.5 & 290 & 34.0 & 300 & 53.0 & 160 & 26.0 & 67 & 12.0 & 25.0 & 6.1 & RMS \\
G333.049+0.032 & 333.049 & +0.032 & 1.3 & 52.0 & 2.9 & 34.0$\dagger$ & 2.6 & 110.0$\phantom{\dagger}$ & 7.9 & 100.0$\phantom{\dagger}$ & 8.0 & 1800 & 120.0 & 1800 & 280.0 & 770 & 150.0 & 280 & 59.0 & 110.0 & 25.0 & RMS \\
G332.978+0.792 & 332.978 & +0.792 & 1.6 & 310.0 & 28.0 & 270.0$\dagger$ & 15.0 & 1100.0$\dagger$ & 20.0 & 940.0$\dagger$ & 16.0 & 12000 & 300.0 & 18000 & 830.0 & 9500 & 460.0 & 3900 & 240.0 & 1600.0 & 110.0 & Known \\
G332.767$-$0.007 & 332.767 & $-$0.007 & 6.4 & 23.0 & 1.1 & 13.0$\phantom{\dagger}$ & 1.1 & 51.0$\phantom{\dagger}$ & 11.0 & 45.0$\phantom{\dagger}$ & 11.0 & 1100 & 190.0 & 1500 & 310.0 & 960 & 140.0 & 450 & 51.0 & 200.0 & 20.0 & RMS \\
G332.148$-$0.446 & 332.148 & $-$0.446 & 7.8 & 330.0 & 26.0 & 240.0$\dagger$ & 21.0 & 960.0$\dagger$ & 30.0 & 820.0$\dagger$ & 15.0 & 13000 & 420.0 & 11000 & 1400.0 & 4000 & 780.0 & 1300 & 300.0 & 510.0 & 130.0 & Known \\
G331.146+0.135 & 331.146 & +0.135 & 2.6 & 3.3 & 0.3 & 2.6$\phantom{\dagger}$ & 0.2 & 10.0$\phantom{\dagger}$ & 0.7 & 12.0$\phantom{\dagger}$ & 0.9 & 270 & 31.0 & 530 & 57.0 & 310 & 21.0 & 140 & 7.0 & 62.0 & 2.9 & RMS \\
G331.093$-$0.130 & 331.093 & $-$0.130 & 3.8 & 1.7 & 0.2 & 1.1$\phantom{\dagger}$ & 0.1 & 17.0$\phantom{\dagger}$ & 1.4 & 14.0$\phantom{\dagger}$ & 1.3 & 190 & 25.0 & 260 & 42.0 & 130 & 17.0 & 52 & 6.2 & 20.0 & 2.4 & RMS \\
G331.089+0.016 & 331.089 & +0.016 & 2.0 & 1.6 & 0.1 & 1.2$\phantom{\dagger}$ & 0.1 & 10.0$\phantom{\dagger}$ & 0.3 & 8.4$\phantom{\dagger}$ & 0.3 & 150 & 26.0 & 300 & 120.0 & 110 & 17.0 & 43 & 6.5 & 17.0 & 2.6 & RMS \\
G330.661+0.579 & 330.661 & +0.579 & 1.8 & 4.0 & 0.3 & 2.7$\phantom{\dagger}$ & 0.2 & 6.3$\phantom{\dagger}$ & 1.2 & 6.8$\phantom{\dagger}$ & 1.3 & 140 & 18.0 & 190 & 81.0 & 97 & 44.0 & 40 & 17.0 & 17.0 & 6.6 & RMS \\
G330.305$-$0.385 & 330.305 & $-$0.385 & 4.3 & 42.0 & 6.6 & 33.0$\dagger$ & 2.9 & 160.0$\phantom{\dagger}$ & 3.1 & 120.0$\dagger$ & 4.0 & 3300 & 100.0 & 3400 & 440.0 & 1300 & 240.0 & 470 & 92.0 & 180.0 & 38.0 & Known \\
G330.292+0.001 & 330.292 & +0.001 & 2.9 & 0.9 & 0.1 & 0.7$\phantom{\dagger}$ & 0.1 & 6.3$\phantom{\dagger}$ & 0.3 & 5.9$\phantom{\dagger}$ & 0.3 & 93 & 1.7 & 200 & 19.0 & 120 & 18.0 & 49 & 9.3 & 20.0 & 4.5 & RMS \\
G329.476+0.841 & 329.476 & +0.841 & 1.3 & 17.0 & 0.5 & 12.0$\phantom{\dagger}$ & 0.4 & 55.0$\phantom{\dagger}$ & 1.0 & 41.0$\dagger$ & 1.1 & 760 & 12.0 & 710 & 20.0 & 280 & 12.0 & 98 & 4.8 & 36.0 & 1.9 & RMS \\
G329.353+0.144 & 329.353 & +0.144 & 1.3 & 45.0 & 1.4 & 45.0$\dagger$ & 1.0 & 220.0$\phantom{\dagger}$ & 5.6 & 210.0$\phantom{\dagger}$ & 3.2 & 7900 & 71.0 & 6800 & 120.0 & 2500 & 92.0 & 960 & 39.0 & 360.0 & 16.0 & Known \\
G328.808+0.633 & 328.808 & +0.633 & 2.2 & 5.5 & 0.1 & 4.3$\dagger$ & 0.1 & 96.0$\dagger$ & 2.1 & 45.0$\dagger$ & 0.3 & 10000 & 14.0 & 6900 & 41.0 & 1700 & 16.0 & 660 & 5.4 & 290.0 & 2.1 & RMS \\
G328.593$-$0.518 & 328.593 & $-$0.518 & 6.7 & 510.0 & 20.0 & 280.0$\dagger$ & 11.0 & 1100.0$\dagger$ & 10.0 & 870.0$\dagger$ & 13.0 & 23000 & 290.0 & 18000 & 1500.0 & 6000 & 940.0 & 2100 & 380.0 & 750.0 & 160.0 & Known \\
G328.310+0.448 & 328.310 & +0.448 & 3.5 & 190.0 & 5.9 & 160.0$\dagger$ & 3.4 & 780.0$\dagger$ & 2.6 & 680.0$\dagger$ & 3.5 & 16000 & 120.0 & 14000 & 400.0 & 5300 & 260.0 & 1900 & 100.0 & 700.0 & 41.0 & Known \\
G328.228$-$0.271 & 328.228 & $-$0.271 & 9.8 & 0.4 & 0.1 & 0.5$\phantom{\dagger}$ & 0.1 & 7.2$\phantom{\dagger}$ & 0.3 & 5.9$\phantom{\dagger}$ & 0.3 & 46 & 4.1 & 130 & 22.0 & 85 & 13.0 & 38 & 5.1 & 16.0 & 2.2 & RMS \\
G328.164+0.587 & 328.164 & +0.587 & 2.5 & 0.6 & 0.1 & 1.0$\phantom{\dagger}$ & 0.1 & 12.0$\phantom{\dagger}$ & 0.4 & 8.9$\phantom{\dagger}$ & 0.5 & 290 & 12.0 & 480 & 32.0 & 260 & 19.0 & 100 & 7.4 & 40.0 & 2.9 & RMS \\
G327.848+0.018 & 327.848 & +0.018 & 1.8 & 1.9 & 0.1 & 1.6$\phantom{\dagger}$ & 0.1 & 8.7$\phantom{\dagger}$ & 0.2 & 8.0$\phantom{\dagger}$ & 0.8 & 150 & 13.0 & 120 & 5.4 & 43 & 5.0 & 15 & 2.4 & 4.5 & 1.2 & RMS \\
G327.356$-$0.101 & 327.356 & $-$0.101 & 1.3 & 3.6 & 1.2 & 2.9$\phantom{\dagger}$ & 0.7 & 5.5$\phantom{\dagger}$ & 0.7 & 6.6$\phantom{\dagger}$ & 0.8 & 190 & 17.0 & 250 & 32.0 & 110 & 14.0 & 40 & 5.1 & 14.0 & 2.1 & RMS \\
G327.313$-$0.536 & 327.313 & $-$0.536 & 1.6 & 1300.0 & 55.0 & 680.0$\dagger$ & 27.0 & 2300.0$\dagger$ & 120.0 & 1900.0$\dagger$ & 130.0 & 77000 & 1900.0 & 64000 & 4400.0 & 21000 & 2500.0 & 7900 & 980.0 & 3000.0 & 400.0 & Known \\
G326.448$-$0.748 & 326.448 & $-$0.748 & 1.8 & 2.9 & 0.1 & 2.2$\phantom{\dagger}$ & 0.1 & 16.0$\phantom{\dagger}$ & 0.5 & 16.0$\phantom{\dagger}$ & 0.3 & 520 & 6.2 & 740 & 13.0 & 380 & 6.7 & 150 & 3.1 & 56.0 & 1.9 & RMS \\
G326.441+0.914 & 326.441 & +0.914 & 8.1 & 280.0 & 5.8 & 170.0$\dagger$ & 4.5 & 670.0$\dagger$ & 10.0 & 520.0$\dagger$ & 9.3 & 12000 & 48.0 & 11000 & 140.0 & 4300 & 63.0 & 1600 & 30.0 & 620.0 & 16.0 & Known \\
G325.516+0.415 & 325.516 & +0.415 & 3.0 & 1.6 & 0.1 & 1.2$\phantom{\dagger}$ & 0.1 & 13.0$\phantom{\dagger}$ & 0.1 & 9.8$\phantom{\dagger}$ & 0.1 & 120 & 2.7 & 220 & 21.0 & 170 & 17.0 & 86 & 7.8 & 38.0 & 3.3 & RMS \\
G324.192+0.109 & 324.192 & +0.109 & 1.1 & 39.0 & 1.1 & 30.0$\dagger$ & 1.1 & 170.0$\phantom{\dagger}$ & 5.6 & 140.0$\dagger$ & 3.8 & 7000 & 51.0 & 6500 & 70.0 & 2500 & 40.0 & 900 & 17.0 & 340.0 & 7.4 & Known \\
G323.741$-$0.255 & 323.741 & $-$0.255 & 1.3 & 45.0 & 1.6 & 36.0$\phantom{\dagger}$ & 0.7 & 130.0$\phantom{\dagger}$ & 1.0 & 100.0$\dagger$ & 1.4 & 3000 & 13.0 & 2900 & 34.0 & 1300 & 20.0 & 560 & 8.8 & 230.0 & 4.2 & RMS \\
G322.407+0.221 & 322.407 & +0.221 & 2.6 & 26.0 & 0.9 & 26.0$\phantom{\dagger}$ & 1.0 & 76.0$\phantom{\dagger}$ & 4.6 & 80.0$\phantom{\dagger}$ & 4.2 & 910 & 52.0 & 1100 & 83.0 & 420 & 44.0 & 140 & 17.0 & 56.0 & 6.7 & Known \\
G322.153+0.613 & 322.153 & +0.613 & 10.8 & 380.0 & 11.0 & 240.0$\dagger$ & 9.1 & 740.0$\dagger$ & 80.0 & 590.0$\dagger$ & 75.0 & 23000 & 480.0 & 25000 & 490.0 & 9600 & 160.0 & 3700 & 55.0 & 1400.0 & 22.0 & Known \\
G320.331$-$0.307 & 320.331 & $-$0.307 & 1.7 & 12.0 & 1.1 & 7.8$\phantom{\dagger}$ & 0.9 & 40.0$\phantom{\dagger}$ & 2.0 & 32.0$\phantom{\dagger}$ & 1.5 & 760 & 26.0 & 750 & 37.0 & 300 & 36.0 & 100 & 17.0 & 36.0 & 7.2 & RMS \\
G320.236+0.417 & 320.236 & +0.417 & 1.3 & 210.0 & 4.4 & 160.0$\phantom{\dagger}$ & 3.7 & 330.0$\phantom{\dagger}$ & 4.0 & 330.0$\phantom{\dagger}$ & 2.1 & 6200 & 200.0 & 8700 & 970.0 & 3200 & 190.0 & 1200 & 78.0 & 500.0 & 31.0 & Known \\
G320.202+0.861 & 320.202 & +0.861 & 1.7 & 29.0 & 1.0 & 17.0$\phantom{\dagger}$ & 0.7 & 65.0$\phantom{\dagger}$ & 1.7 & 45.0$\dagger$ & 1.0 & 1100 & 32.0 & 1000 & 96.0 & 420 & 74.0 & 160 & 34.0 & 65.0 & 15.0 & RMS \\
G320.153+0.780 & 320.153 & +0.780 & 1.6 & 260.0 & 16.0 & 240.0$\dagger$ & 8.8 & 790.0$\phantom{\dagger}$ & 9.0 & 740.0$\phantom{\dagger}$ & 9.4 & 8800 & 200.0 & 9800 & 350.0 & 4000 & 140.0 & 1500 & 59.0 & 580.0 & 25.0 & Known \\
G319.874+0.770 & 319.874 & +0.770 & 1.8 & 170.0 & 15.0 & 130.0$\phantom{\dagger}$ & 9.2 & 260.0$\phantom{\dagger}$ & 14.0 & 270.0$\phantom{\dagger}$ & 13.0 & 4600 & 260.0 & 5300 & 690.0 & 2200 & 350.0 & 850 & 140.0 & 350.0 & 59.0 & Known \\
G318.914$-$0.164 & 318.914 & $-$0.164 & 2.0 & 16.0 & 2.8 & 15.0$\dagger$ & 2.2 & 100.0$\phantom{\dagger}$ & 2.2 & 72.0$\dagger$ & 2.1 & 2200 & 34.0 & 2000 & 110.0 & 820 & 72.0 & 290 & 38.0 & 110.0 & 21.0 & RMS \\
G318.725$-$0.223 & 318.725 & $-$0.223 & 1.5 & 5.2 & 0.1 & 3.7$\phantom{\dagger}$ & 0.1 & 14.0$\phantom{\dagger}$ & 0.1 & 15.0$\phantom{\dagger}$ & 0.1 & 330 & 3.3 & 410 & 11.0 & 170 & 8.2 & 62 & 4.5 & 23.0 & 2.7 & RMS \\
G318.050+0.086 & 318.050 & +0.086 & 2.7 & 2.1 & 0.1 & 2.7$\phantom{\dagger}$ & 0.1 & 33.0$\phantom{\dagger}$ & 0.1 & 28.0$\phantom{\dagger}$ & 0.1 & 1900 & 3.8 & 2000 & 8.7 & 860 & 9.9 & 310 & 6.0 & 100.0 & 3.2 & RMS \\
G317.889$-$0.058 & 317.889 & $-$0.058 & 1.3 & 5.9 & 1.3 & 5.2$\phantom{\dagger}$ & 0.7 & 24.0$\phantom{\dagger}$ & 0.4 & 21.0$\phantom{\dagger}$ & 0.7 & 550 & 65.0 & 720 & 310.0 & 230 & 20.0 & 86 & 8.7 & 35.0 & 3.9 & RMS \\
G317.748+0.011 & 317.748 & +0.011 & 3.6 & 3.0 & 0.5 & 2.3$\phantom{\dagger}$ & 0.5 & 6.1$\phantom{\dagger}$ & 0.6 & 7.1$\phantom{\dagger}$ & 0.7 & 180 & 6.1 & 250 & 24.0 & 120 & 19.0 & 48 & 10.0 & 20.0 & 5.2 & RMS \\
G316.156$-$0.492 & 316.156 & $-$0.492 & 6.2 & 34.0 & 1.2 & 20.0$\dagger$ & 0.2 & 110.0$\phantom{\dagger}$ & 0.7 & 72.0$\dagger$ & 2.1 & 2400 & 16.0 & 2400 & 62.0 & 1100 & 35.0 & 410 & 14.0 & 160.0 & 5.6 & Known \\
G315.312$-$0.273 & 315.312 & $-$0.273 & 7.6 & 12.0 & 1.2 & 9.4$\phantom{\dagger}$ & 0.6 & 23.0$\phantom{\dagger}$ & 1.2 & 24.0$\phantom{\dagger}$ & 1.2 & 540 & 13.0 & 950 & 93.0 & 480 & 70.0 & 210 & 32.0 & 95.0 & 15.0 & Known \\
G313.458+0.191 & 313.458 & +0.191 & 2.3 & 15.0 & 0.4 & 14.0$\phantom{\dagger}$ & 0.5 & 85.0$\phantom{\dagger}$ & 2.1 & 71.0$\phantom{\dagger}$ & 3.3 & 1600 & 40.0 & 1800 & 29.0 & 740 & 13.0 & 270 & 6.3 & 110.0 & 2.8 & RMS \\
G313.446+0.176 & 313.446 & +0.176 & 4.3 & 20.0 & 3.8 & 18.0$\phantom{\dagger}$ & 1.5 & 84.0$\phantom{\dagger}$ & 4.5 & 66.0$\phantom{\dagger}$ & 6.7 & 1600 & 120.0 & 1900 & 250.0 & 820 & 120.0 & 320 & 47.0 & 130.0 & 20.0 & Known \\
G312.953$-$0.449 & 312.953 & $-$0.449 & 4.0 & 65.0 & 9.2 & 44.0$\phantom{\dagger}$ & 2.8 & 73.0$\phantom{\dagger}$ & 6.2 & 74.0$\phantom{\dagger}$ & 6.3 & 1300 & 180.0 & 3300 & 150.0 & 1700 & 76.0 & 640 & 50.0 & 270.0 & 28.0 & Known \\
G312.548$-$0.281 & 312.548 & $-$0.281 & 5.1 & 1.9 & 0.3 & 1.3$\phantom{\dagger}$ & 0.1 & 9.3$\phantom{\dagger}$ & 0.1 & 8.5$\phantom{\dagger}$ & 0.1 & 180 & 4.0 & 300 & 27.0 & 160 & 17.0 & 66 & 7.3 & 25.0 & 3.1 & RMS \\
G312.383$-$0.415 & 312.383 & $-$0.415 & 2.2 & 4.0 & 0.2 & 2.7$\phantom{\dagger}$ & 0.2 & 23.0$\phantom{\dagger}$ & 0.7 & 21.0$\phantom{\dagger}$ & 0.7 & 320 & 7.1 & 410 & 36.0 & 240 & 26.0 & 100 & 12.0 & 45.0 & 5.3 & RMS \\
G312.112+0.314 & 312.112 & +0.314 & 3.0 & 33.0 & 1.7 & 20.0$\dagger$ & 1.3 & 120.0$\phantom{\dagger}$ & 1.5 & 92.0$\dagger$ & 1.9 & 2900 & 28.0 & 2900 & 73.0 & 1300 & 36.0 & 510 & 15.0 & 200.0 & 6.8 & Known \\
G311.627+0.270 & 311.627 & +0.270 & 2.2 & 20.0 & 1.7 & 21.0$\phantom{\dagger}$ & 12.0 & 110.0$\phantom{\dagger}$ & 15.0 & 71.0$\phantom{\dagger}$ & 2.8 & 3900 & 59.0 & 4400 & 190.0 & 2000 & 120.0 & 770 & 55.0 & 300.0 & 26.0 & Known \\
G311.426+0.596 & 311.426 & +0.596 & 4.2 & 14.0 & 0.2 & 8.9$\phantom{\dagger}$ & 0.2 & 30.0$\phantom{\dagger}$ & 0.4 & 26.0$\phantom{\dagger}$ & 0.6 & 540 & 11.0 & 420 & 13.0 & 140 & 12.0 & 45 & 5.8 & 15.0 & 2.7 & RMS \\
G310.994+0.389 & 310.994 & +0.389 & 15.3 & 240.0 & 9.0 & 320.0$\phantom{\dagger}$ & 160.0 & 240.0$\phantom{\dagger}$ & 17.0 & 210.0$\phantom{\dagger}$ & 47.0 & 3900 & 210.0 & 6800 & 260.0 & 2900 & 280.0 & 1100 & 140.0 & 440.0 & 66.0 & Known \\
G310.142+0.758 & 310.142 & +0.758 & 7.8 & 2.7 & 0.1 & 2.1$\phantom{\dagger}$ & 0.2 & 25.0$\phantom{\dagger}$ & 0.1 & 21.0$\phantom{\dagger}$ & 0.2 & 1600 & 3.8 & 1600 & 7.3 & 690 & 6.2 & 250 & 3.2 & 81.0 & 1.4 & RMS \\
G309.921+0.479 & 309.921 & +0.479 & 3.1 & 23.0 & 0.7 & 13.0$\phantom{\dagger}$ & 0.4 & 140.0$\phantom{\dagger}$ & 7.7 & 140.0$\dagger$ & 7.3 & 4500 & 130.0 & 4400 & 290.0 & 1900 & 140.0 & 730 & 54.0 & 290.0 & 23.0 & RMS \\
G309.548$-$0.737 & 309.548 & $-$0.737 & 4.4 & 56.0 & 3.1 & 60.0$\phantom{\dagger}$ & 12.0 & 210.0$\phantom{\dagger}$ & 14.0 & 230.0$\phantom{\dagger}$ & 1.1 & 2000 & 30.0 & 2100 & 150.0 & 890 & 96.0 & 350 & 41.0 & 150.0 & 18.0 & Known \\
G309.177$-$0.028 & 309.177 & $-$0.028 & 1.9 & 7.8 & 0.5 & 6.6$\phantom{\dagger}$ & 0.5 & 26.0$\phantom{\dagger}$ & 0.6 & 25.0$\phantom{\dagger}$ & 0.5 & 450 & 2.4 & 510 & 29.0 & 250 & 21.0 & 100 & 9.0 & 40.0 & 3.6 & RMS \\
G308.918+0.122 & 308.918 & +0.122 & 3.7 & 74.0 & 1.5 & 47.0$\dagger$ & 0.5 & 130.0$\phantom{\dagger}$ & 1.7 & 130.0$\dagger$ & 1.3 & 3400 & 14.0 & 2900 & 61.0 & 1100 & 36.0 & 380 & 14.0 & 140.0 & 6.6 & RMS \\
G308.647+0.579 & 308.647 & +0.579 & 2.4 & 730.0 & 40.0 & 660.0$\phantom{\dagger}$ & 31.0 & 3100.0$\phantom{\dagger}$ & 67.0 & 3100.0$\phantom{\dagger}$ & 59.0 & 24000 & 910.0 & 27000 & 1500.0 & 10000 & 610.0 & 3700 & 210.0 & 1500.0 & 92.0 & Known \\
G308.201$-$1.020 & 308.201 & $-$1.020 & 1.9 & 1.7 & 0.2 & 1.8$\phantom{\dagger}$ & 0.1 & 16.0$\phantom{\dagger}$ & 0.2 & 12.0$\phantom{\dagger}$ & 0.1 & 110 & 1.5 & 110 & 5.7 & 59 & 4.2 & 24 & 1.9 & 10.0 & 1.0 & RMS \\
G307.736$-$0.594 & 307.736 & $-$0.594 & 2.8 & 0.6 & 0.1 & 0.6$\phantom{\dagger}$ & 0.1 & 17.0$\phantom{\dagger}$ & 4.3 & 14.0$\phantom{\dagger}$ & 6.4 & 97 & 18.0 & 240 & 98.0 & 140 & 55.0 & 66 & 25.0 & 30.0 & 11.0 & RMS \\
G307.620$-$0.320 & 307.620 & $-$0.320 & 25.9 & 260.0 & 34.0 & 260.0$\phantom{\dagger}$ & 16.0 & 1400.0$\phantom{\dagger}$ & 15.0 & 1400.0$\phantom{\dagger}$ & 24.0 & 11000 & 420.0 & 10000 & 2000.0 & 3600 & 1300.0 & 1200 & 550.0 & 490.0 & 240.0 & Known \\
G307.614$-$0.256 & 307.614 & $-$0.256 & 8.6 & 77.0 & 1.6 & 65.0$\phantom{\dagger}$ & 4.5 & 310.0$\phantom{\dagger}$ & 47.0 & 250.0$\dagger$ & 53.0 & 3100 & 160.0 & 3000 & 63.0 & 1200 & 37.0 & 440 & 18.0 & 170.0 & 8.2 & RMS \\
G307.561$-$0.587 & 307.561 & $-$0.587 & 3.4 & 20.0 & 2.7 & 15.0$\phantom{\dagger}$ & 0.4 & 98.0$\phantom{\dagger}$ & 1.4 & 54.0$\dagger$ & 2.4 & 1900 & 24.0 & 2500 & 170.0 & 1300 & 130.0 & 530 & 63.0 & 220.0 & 29.0 & RMS \\
G306.315$-$0.361 & 306.315 & $-$0.361 & 9.4 & 19.0 & 0.6 & 21.0$\phantom{\dagger}$ & 1.0 & 93.0$\phantom{\dagger}$ & 0.7 & 100.0$\phantom{\dagger}$ & 0.6 & 1000 & 6.2 & 1100 & 71.0 & 450 & 66.0 & 170 & 35.0 & 72.0 & 18.0 & Known \\
G305.603$-$0.690 & 305.603 & $-$0.690 & 3.2 & 1.6 & 0.3 & 1.4$\phantom{\dagger}$ & 0.1 & 12.0$\phantom{\dagger}$ & 0.1 & 9.1$\phantom{\dagger}$ & 0.2 & 67 & 3.2 & 110 & 12.0 & 76 & 7.1 & 37 & 3.3 & 16.0 & 1.6 & RMS \\
G303.115$-$0.947 & 303.115 & $-$0.947 & 13.2 & 9.9 & 0.7 & 11.0$\phantom{\dagger}$ & 0.6 & 72.0$\phantom{\dagger}$ & 0.6 & 41.0$\dagger$ & 12.0 & 560 & 22.0 & 470 & 30.0 & 180 & 15.0 & 64 & 6.1 & 26.0 & 2.6 & Known \\
G302.690+0.190 & 302.690 & +0.190 & 7.4 & 120.0 & 13.0 & 90.0$\phantom{\dagger}$ & 6.3 & 300.0$\phantom{\dagger}$ & 5.2 & 300.0$\phantom{\dagger}$ & 8.9 & 4000 & 180.0 & 5700 & 550.0 & 2200 & 240.0 & 780 & 86.0 & 310.0 & 36.0 & Known \\
G302.152$-$0.949 & 302.152 & $-$0.949 & 1.2 & 5.8 & 0.4 & 5.5$\phantom{\dagger}$ & 0.3 & 32.0$\phantom{\dagger}$ & 0.5 & 26.0$\phantom{\dagger}$ & 0.7 & 350 & 4.2 & 300 & 15.0 & 130 & 7.5 & 49 & 3.0 & 20.0 & 1.3 & RMS \\
G302.025$-$0.044 & 302.025 & $-$0.044 & 2.3 & 14.0 & 1.8 & 11.0$\phantom{\dagger}$ & 1.0 & 97.0$\phantom{\dagger}$ & 1.1 & 52.0$\dagger$ & 1.2 & 1700 & 13.0 & 1700 & 78.0 & 750 & 51.0 & 270 & 23.0 & 100.0 & 10.0 & Known \\
G298.228$-$0.331 & 298.228 & $-$0.331 & 2.7 & 340.0 & 15.0 & 300.0$\dagger$ & 16.0 & 1500.0$\dagger$ & 80.0 & 1300.0$\dagger$ & 45.0 & 23000 & 760.0 & 14000 & 900.0 & 4300 & 250.0 & 2100 & 120.0 & 540.0 & 31.0 & Known \\
G298.187$-$0.782 & 298.187 & $-$0.782 & 6.3 & 46.0 & 1.0 & 36.0$\dagger$ & 0.5 & 150.0$\phantom{\dagger}$ & 2.3 & 120.0$\dagger$ & 1.7 & 3000 & 21.0 & 2200 & 26.0 & 760 & 21.0 & 370 & 14.0 & 97.0 & 4.7 & Known \\

\end{longtable}
\tablebib{(HRDS)~\citet{anderson11a}; (Known)~\hiismall\ regions known prior to the HRDS described in \citet{anderson11a}; (RMS)~\citet{urquhart08}.}
\end{tiny}
\end{landscape}
}

\clearpage
\renewcommand{\tabcolsep}{2.5pt}
\longtab{2}{
\begin{landscape}
\begin{tiny}
\begin{longtable}{llrrrrrrrrrrrrrrrrrrrrrl}
\caption{\label{tab:pn}The fluxes of PNe from 8.0\,\micron\ to 500\,\micron.}\\
\hline\hline \\ [-2ex]
Gal. Name & Name & $\ell$\phantom{00} & $b$\phantom{00} & Size & $F_{8}$\,\, & $\sigma_{8}$ & $F_{12}$\,\, & $\sigma_{12}$ & $F_{22}$\,\, & $\sigma_{22}$ & $F_{24}$\,\, & $\sigma_{24}$ & $F_{70}$\,\, & $\sigma_{70}$ & $F_{160}$\, & $\sigma_{160}$ & $F_{250}$\, & $\sigma_{250}$ & $F_{350}$\, & $\sigma_{350}$ & $F_{500}$ & $\sigma_{500}$ & Reference\\
& & deg.\,               & deg.\,            & $\arcsec$   & Jy\,\,     & Jy\,\,      & Jy\,\,      & Jy\,\,      & Jy\,\,      & Jy\,\,      & Jy\,\,      & Jy\,\,       & Jy\,\,       & Jy\,\,       & Jy\,\,       & Jy\,\,     & Jy\,\,      & Jy\,\,      & Jy\,\,       & Jy\,\,       & Jy\,\,       & Jy\,\, &           \\
\hline \\ [-2ex]
\endfirsthead
\caption{continued.}\\
\hline\hline \\ [-2ex]
Gal. Name & $\ell$\phantom{00} & $b$\phantom{00} & Size & $F_{8}$\,\, & $\sigma_{8}$ & $F_{12}$\,\, & $\sigma_{12}$ & $F_{22}$\,\, & $\sigma_{22}$ & $F_{24}$\,\, & $\sigma_{24}$ & $F_{70}$\,\, & $\sigma_{70}$ & $F_{160}$\, & $\sigma_{160}$ & $F_{250}$\, & $\sigma_{250}$ & $F_{350}$\, & $\sigma_{350}$ & $F_{500}$\, & $\sigma_{500}$ & Reference\\
& & deg.\,               & deg.\,            & $\arcsec$   & Jy\,\,     & Jy\,\,      & Jy\,\,      & Jy\,\,      & Jy\,\,      & Jy\,\,      & Jy\,\,      & Jy\,\,       & Jy\,\,       & Jy\,\,       & Jy\,\,       & Jy\,\,     & Jy\,\,      & Jy\,\,      & Jy\,\,       & Jy\,\,       & Jy\,\,       & Jy\,\, &           \\
\hline
\endhead
\hline
\endfoot
PNG002.2+00.5 & Terz N 2337 & 2.251 & +0.552 &           27 & 0.190 & 0.015 & 0.570 & 0.038 & 2.10 & 0.12 & 2.30$\phantom{\dagger}$ & 0.63 & 6.50 & 0.42 & 8.40 & 0.93 & 2.70 & 1.30 & 1.60 & 0.53 & 0.77 & 0.31 & ESO \\
PNG001.8$-$00.5 & JaSt 81 & 1.868 & $-$0.534 &           32 & 0.089 & 0.005 & 0.270 & 0.031 & 1.60 & 0.09 & 1.80$\phantom{\dagger}$ & 0.49 & 3.90 & 0.59 & 2.00 & 1.70 & 1.20 & 0.38 & \nodata & \nodata & \nodata & \nodata & ESO \\
PNG001.6+00.1 & JaSt 63 & 1.650 & +0.188 &           12 & \nodata & \nodata & \nodata & \nodata & \nodata & \nodata & 0.06$\phantom{\dagger}$ & 0.02 & 0.97 & 1.50 & \nodata & \nodata & \nodata & \nodata & \nodata & \nodata & \nodata & \nodata & ESO \\
PNG001.6+00.9 & JaSt 55 & 1.635 & +0.997 &           18 & 0.023 & 0.001 & 0.150 & 0.029 & 0.41 & 0.02 & 0.46$\phantom{\dagger}$ & 0.12 & 2.50 & 0.12 & 0.89 & 0.11 & \nodata & \nodata & \nodata & \nodata & \nodata & \nodata & ESO \\
PNG001.5$-$00.7 & K 6$-$17 & 1.509 & $-$0.766 &           36 & 0.250 & 0.015 & 0.690 & 0.130 & 5.90 & 0.17 & 5.10$\phantom{\dagger}$ & 1.40 & 5.40 & 1.10 & 1.10 & 0.43 & \nodata & \nodata & \nodata & \nodata & \nodata & \nodata & ESO \\
PNG001.2+00.7 & JaSt 56 & 1.235 & +0.712 &           18 & 0.011 & 0.002 & 0.071 & 0.010 & 0.37 & 0.09 & 0.35$\phantom{\dagger}$ & 0.10 & 0.63 & 0.41 & \nodata & \nodata & \nodata & \nodata & \nodata & \nodata & \nodata & \nodata & ESO \\
PNG001.1+00.8 & JaSt 54 & 1.138 & +0.808 &           30 & 1.200 & 0.030 & 1.300 & 0.120 & 7.60 & 0.17 & 7.90$\phantom{\dagger}$ & 2.10 & 67.00 & 0.98 & 46.00 & 4.80 & 22.00 & 1.30 & 7.20 & 1.80 & 3.50 & 0.45 & ESO \\
PNG000.8$-$00.6 & JaSt 71 & 0.869 & $-$0.696 &           13 & 0.052 & 0.004 & 0.400 & 0.034 & 2.90 & 0.21 & 2.70$\phantom{\dagger}$ & 0.73 & 3.20 & 0.21 & 0.99 & 0.32 & \nodata & \nodata & \nodata & \nodata & \nodata & \nodata & ESO \\
PNG000.7$-$00.8 & JaSt 74 & 0.748 & $-$0.869 &           17 & 0.014 & 0.003 & 0.150 & 0.028 & 0.66 & 0.03 & 0.72$\phantom{\dagger}$ & 0.20 & 0.82 & 0.19 & \nodata & \nodata & \nodata & \nodata & \nodata & \nodata & \nodata & \nodata & ESO \\
PNG000.6$-$01.0 & JaSt 77 & 0.625 & $-$1.046 &           32 & 0.150 & 0.004 & 1.000 & 0.035 & 5.70 & 0.08 & 4.90$\phantom{\dagger}$ & 1.30 & 0.84 & 0.17 & \nodata & \nodata & \nodata & \nodata & \nodata & \nodata & \nodata & \nodata & ESO \\
PNG000.1$-$01.0 & JaSt 69 & 0.187 & $-$1.047 &           16 & 0.011 & 0.001 & 0.042 & 0.007 & 0.07 & 0.04 & 0.09$\phantom{\dagger}$ & 0.03 & 0.11 & 0.16 & \nodata & \nodata & \nodata & \nodata & \nodata & \nodata & \nodata & \nodata & ESO \\
PNG359.3$-$00.9 & Hb 5 & 359.355 & $-$0.984 &           43 & 5.900 & 0.027 & 2.400 & 0.083 & 44.00 & 6.50 & 26.00$\dagger$ & 7.00 & 100.00 & 1.10 & 38.00 & 2.20 & 18.00 & 1.70 & 7.70 & 1.10 & 3.50 & 0.51 & ESO \\
PNG358.9$-$00.7 & M 1$-$26 & 358.960 & $-$0.721 &           54 & 2.600 & 0.016 & 6.200 & 0.250 & 83.00 & 3.50 & 31.00$\dagger$ & 8.50 & 36.00 & 2.70 & 7.90 & 2.70 & \nodata & \nodata & \nodata & \nodata & \nodata & \nodata & ESO \\
PNG358.6$-$01.1 & JaSt 58 & 358.695 & $-$1.111 &           15 & 0.057 & 0.003 & 0.260 & 0.048 & 0.42 & 0.14 & 0.38$\phantom{\dagger}$ & 0.12 & 4.30 & 0.66 & 2.10 & 1.30 & 1.10 & 0.37 & \nodata & \nodata & \nodata & \nodata & ESO \\
PNG358.6+00.7 & JaSt 16 & 358.637 & +0.756 &           32 & 0.030 & 0.003 & 0.220 & 0.006 & 1.80 & 0.09 & 2.10$\phantom{\dagger}$ & 0.03 & 4.70 & 0.70 & 5.70 & 1.20 & \nodata & \nodata & \nodata & \nodata & \nodata & \nodata & ESO \\
PNG356.9+00.9 & PPA1734$-$3102 & 356.951 & +0.911 &           44 & 0.200 & 0.003 & 3.500 & 0.210 & 5.20 & 0.15 & 4.30$\phantom{\dagger}$ & 0.05 & 6.00 & 0.17 & 1.90 & 0.58 & \nodata & \nodata & \nodata & \nodata & \nodata & \nodata & Mash I \\
PNG347.2$-$00.8 & PHR 1714$-$4006 & 347.199 & $-$0.869 &           19 & 0.016 & 0.006 & 0.050 & 0.019 & 0.04 & 0.01 & 0.07$\phantom{\dagger}$ & 0.01 & 0.68 & 0.14 & 0.87 & 0.40 & 0.69 & 0.28 & 0.27 & 0.13 & \nodata & \nodata & Mash I \\
PNG345.4+00.1 & IC4637 & 345.479 & +0.140 &           31 & 0.230 & 0.026 & 0.920 & 0.052 & 13.00 & 0.10 & 13.00$\phantom{\dagger}$ & 0.09 & 7.90 & 0.35 & \nodata & \nodata & \nodata & \nodata & \nodata & \nodata & \nodata & \nodata & ESO \\
PNG343.9+00.8 & H 1$-$5 & 343.992 & +0.836 &           66 & 0.710 & 0.003 & 3.300 & 0.270 & 25.00 & 0.54 & 19.00$\phantom{\dagger}$ & 0.46 & 25.00 & 0.32 & 6.50 & 2.10 & 2.40 & 1.10 & 1.70 & 0.93 & 0.83 & 0.74 & ESO \\
PNG342.7+00.7 & H 1$-$3 & 342.746 & +0.752 &           23 & 0.055 & 0.006 & 0.220 & 0.006 & 0.75 & 0.07 & 0.63$\phantom{\dagger}$ & 0.02 & 6.20 & 0.31 & 2.70 & 0.55 & 1.60 & 0.51 & 0.58 & 0.19 & 0.25 & 0.08 & ESO \\
PNG342.2$-$00.3 & PM 1$-$119 & 342.226 & $-$0.380 &           61 & 0.200 & 0.011 & 1.700 & 0.080 & 17.00 & 0.67 & 12.00$\phantom{\dagger}$ & 0.31 & 13.00 & 0.90 & 2.20 & 1.70 & 1.20 & 0.74 & \nodata & \nodata & \nodata & \nodata & ESO \\
PNG339.1+00.9 & PHR 1639$-$4516 & 339.099 & +0.989 &           18 & \nodata & \nodata & 0.036 & 0.008 & 0.04 & 0.01 & 0.06$\phantom{\dagger}$ & 0.01 & 0.45 & 0.13 & \nodata & \nodata & \nodata & \nodata & \nodata & \nodata & \nodata & \nodata & Mash I \\
PNG337.3+00.6 & PHR 1633$-$4650 & 337.314 & +0.636 &           20 & 0.100 & 0.006 & 0.310 & 0.310 & 1.30 & 0.12 & 1.50$\phantom{\dagger}$ & 0.11 & 18.00 & 0.56 & 15.00 & 2.20 & 7.10 & 1.50 & 2.60 & 1.30 & 1.10 & 0.69 & Mash I \\
PNG333.9+00.6 & PHR 1619$-$4914 & 333.929 & +0.686 &           37 & 0.790 & 0.072 & 4.900 & 0.270 & 21.00 & 0.36 & 19.00$\phantom{\dagger}$ & 0.26 & 6.50 & 0.41 & \nodata & \nodata & \nodata & \nodata & \nodata & \nodata & \nodata & \nodata & Mash I \\
PNG329.5$-$00.8 & MPA 1605$-$5319 & 329.523 & $-$0.796 &           22 & 0.045 & 0.005 & 0.150 & 0.039 & 0.46 & 0.04 & 0.53$\phantom{\dagger}$ & 0.02 & 2.40 & 0.11 & 0.87 & 0.01 & 0.29 & 0.06 & 0.10 & 0.10 & \nodata & \nodata & Mash II \\
PNG328.3+00.7 & PHR1552$-$5254 & 328.357 & +0.766 &           19 & 0.075 & 0.002 & 0.290 & 0.095 & 0.40 & 0.07 & 0.48$\phantom{\dagger}$ & 0.02 & 5.10 & 0.42 & \nodata & \nodata & \nodata & \nodata & \nodata & \nodata & \nodata & \nodata & Mash I \\
PNG322.4$-$00.1 & Pe2$-$8 & 322.469 & $-$0.175 &           51 & 3.700 & 0.063 & 6.000 & 1.700 & 39.00 & 1.40 & 23.00$\phantom{\dagger}$ & 0.28 & 22.00 & 0.18 & 5.50 & 0.94 & 2.00 & 0.38 & 1.00 & 0.11 & 0.60 & 0.16 & ESO \\
PNG322.4$-$00.1a & MPA 1523$-$5710 & 322.418 & $-$0.172 &           26 & 0.040 & 0.003 & 0.190 & 0.027 & 0.78 & 0.21 & 0.71$\phantom{\dagger}$ & 0.10 & 4.10 & 0.36 & 3.40 & 0.59 & 1.40 & 0.38 & 0.58 & 0.18 & \nodata & \nodata & Mash II \\
PNG322.2$-$00.7 & PM 1$-$90 & 322.207 & $-$0.743 &           54 & 0.180 & 0.010 & 0.650 & 0.044 & 4.30 & 0.13 & 4.30$\phantom{\dagger}$ & 0.10 & 6.00 & 0.30 & 1.60 & 0.26 & 0.58 & 0.18 & \nodata & \nodata & \nodata & \nodata & Mash II \\
PNG322.2$-$00.4 & BMP 1522$-$5729 & 322.198 & $-$0.410 &           36 & 0.100 & 0.009 & 0.440 & 0.048 & 1.70 & 0.08 & 1.70$\phantom{\dagger}$ & 0.08 & 8.40 & 0.18 & 4.20 & 1.10 & 1.60 & 1.70 & 0.68 & 0.17 & \nodata & \nodata & Mash II \\
PNG318.9+00.6 & PHR 1457$-$5812 & 318.931 & +0.695 &           33 & 0.130 & 0.008 & 0.620 & 0.170 & 1.70 & 0.08 & 1.90$\phantom{\dagger}$ & 0.02 & 8.10 & 0.27 & 4.20 & 1.40 & \nodata & \nodata & \nodata & \nodata & \nodata & \nodata & Mash I \\
PNG316.2+00.8 & GLMP387 & 316.247 & +0.884 &           38 & 0.700 & 0.007 & 2.400 & 0.230 & 15.00 & 0.27 & 11.00$\phantom{\dagger}$ & 0.14 & 14.00 & 0.11 & 4.30 & 0.80 & \nodata & \nodata & \nodata & \nodata & \nodata & \nodata & ESO \\
PNG315.9+00.3 & PHR 1437$-$5949 & 315.945 & +0.331 &           51 & \nodata & \nodata & \nodata & \nodata & 0.37 & 0.08 & 0.54$\phantom{\dagger}$ & 0.08 & 0.65 & 0.23 & \nodata & \nodata & \nodata & \nodata & \nodata & \nodata & \nodata & \nodata & Mash I \\
PNG315.0$-$00.3 & Hen 2$-$111 & 315.030 & $-$0.373 &           41 & 0.230 & 0.010 & 1.200 & 0.094 & 3.30 & 0.10 & 3.80$\phantom{\dagger}$ & 0.06 & 11.00 & 0.29 & 13.00 & 2.60 & 5.30 & 4.90 & 2.10 & 0.96 & 1.10 & 0.53 & ESO \\
PNG313.3+00.3 & \, & 313.355 & +0.312 &           32 & 0.130 & 0.006 & 0.250 & 0.092 & 0.85 & 0.10 & 0.92$\phantom{\dagger}$ & 0.06 & 0.81 & 0.22 & \nodata & \nodata & \nodata & \nodata & \nodata & \nodata & \nodata & \nodata & Cohen \\
PNG309.0+00.8 & Hen 2$-$96 & 309.020 & +0.890 &           38 & 0.660 & 0.006 & 1.300 & 0.170 & 12.00 & 0.17 & 7.70$\phantom{\dagger}$ & 0.11 & 16.00 & 0.27 & 5.60 & 0.61 & 2.40 & 0.28 & 1.30 & 0.26 & 0.51 & 0.19 & ESO \\
PNG308.1$-$00.5 & MPA 1337$-$6258 & 308.182 & $-$0.585 &           23 & \nodata & \nodata & 0.046 & 0.020 & 0.08 & 0.02 & 0.11$\phantom{\dagger}$ & 0.01 & 1.20 & 0.09 & 0.68 & 0.24 & \nodata & \nodata & \nodata & \nodata & \nodata & \nodata & Mash II \\
PNG306.4$-$00.6 & Th 2$-$A & 306.416 & $-$0.688 &           27 & 0.260 & 0.009 & 0.500 & 0.014 & 3.10 & 0.15 & 2.80$\phantom{\dagger}$ & 0.02 & 4.10 & 0.10 & 2.50 & 0.53 & 0.86 & 0.47 & 0.38 & 0.17 & 0.23 & 0.07 & ESO \\
PNG305.6$-$00.9 & MPA 1315$-$6338 & 305.599 & $-$0.898 &           34 & 0.030 & 0.003 & 0.092 & 0.017 & 0.48 & 0.07 & 0.58$\phantom{\dagger}$ & 0.06 & 1.50 & 0.08 & 0.69 & 0.31 & \nodata & \nodata & \nodata & \nodata & \nodata & \nodata & Mash II \\
PNG302.3$-$00.5 & PHR1246$-$6324 & 302.373 & $-$0.541 &           28 & 0.140 & 0.006 & 0.250 & 0.075 & 0.88 & 0.07 & 0.89$\phantom{\dagger}$ & 0.01 & 5.60 & 0.28 & 3.10 & 0.51 & 1.20 & 0.34 & 0.50 & 0.11 & 0.23 & 0.07 & Mash I \\
PNG301.1$-$00.4 & MPA 1235$-$6318 & 301.127 & $-$0.485 &           21 & 0.008 & 0.001 & 0.028 & 0.002 & 0.12 & 0.03 & 0.13$\phantom{\dagger}$ & 0.01 & 0.11 & 0.08 & \nodata & \nodata & \nodata & \nodata & \nodata & \nodata & \nodata & \nodata & Mash II \\
PNG300.4$-$00.9 & Hen 2$-$84 & 300.429 & $-$0.982 &           36 & 0.031 & 0.001 & 0.140 & 0.030 & 0.34 & 0.02 & 0.36$\phantom{\dagger}$ & 0.01 & 2.40 & 0.06 & 2.10 & 0.28 & 1.10 & 0.30 & 0.74 & 0.07 & 0.43 & 0.15 & ESO \\
PNG296.7$-$00.2 & MPA1157$-$6226 & 296.771 & $-$0.211 &           18 & 0.003 & 0.001 & \nodata & \nodata & 0.11 & 0.03 & 0.11$\phantom{\dagger}$ & 0.01 & 1.20 & 0.08 & 0.77 & 0.36 & 0.44 & 0.16 & \nodata & \nodata & \nodata & \nodata & Mash II \\

\end{longtable}
\tablebib{``ESO'' for \citet{kohoutek01}, ``MASH~I'' for \citet{parker06}, ``MASH~II'' for
  \citet{miszalski08}, and ``Cohen'' for \citet{cohen05}.}
\end{tiny}
\end{landscape}
}

\end{document}